\documentclass{aa} 

\usepackage{natbib}
\usepackage{multirow}
\usepackage{hyperref}
\bibpunct{(}{)}{;}{a}{}{,} 

\usepackage{graphicx}
\usepackage{adjustbox}
\usepackage{tabularx}
\usepackage[varg]{txfonts}

\begin{document} 

\defcitealias{Mattila2018}{M18}
\defcitealias{Kool2020}{K20}

   \title{Energetic nuclear transients in luminous and ultraluminous infrared galaxies}
   \titlerunning{Energetic nuclear transients in U/LIRGs}

   \author{T. M. Reynolds\inst{1,2,3} \and 
          S. Mattila\inst{1} \and 
         A. Efstathiou\inst{4} \and
         E. Kankare\inst{1}  \and
         E. Kool\inst{5} \and
         S. Ryder\inst{6,7} \and
         L. Pe\~{n}a-Mo\~{n}ino\inst{8} \and
         M. A. P\'{e}rez-Torres\inst{8}}

   \institute{Tuorla observatory, Department of Physics and Astronomy, University of Turku, FI-20014 Turku, Finland 
     \and
   Cosmic Dawn Center (DAWN), Denmark  
   \and
   Niels Bohr Institute, University of Copenhagen, Jagtvej 128, 2200 København N, Denmark
         \and
             School of Sciences, European University Cyprus, Diogenes Street, Engomi, 1516 Nicosia, Cyprus
         \and
             The Oskar Klein Centre, Department of Astronomy, Stockholm University, AlbaNova, SE-10691, Stockholm, Sweden
         \and
            School of Mathematical and Physical Sciences, Macquarie University, NSW 2109, Sydney, Australia
         \and
            Astronomy, Astrophysics and Astrophotonics Research Centre, Macquarie University, Sydney, NSW 2109, Australia
         \and
             Instituto de Astrof\'{i}sica de Andaluc\'{i}a (CSIC), Glorieta de la Astronom\'{i}a s/n, E-18080 Granada, Spain
             }

    \date{Submitted to Astronomy \& Astrophysics 9 February 2022 / Accepted 24 May 2022}

\abstract{Energetic nuclear outbursts have been discovered in luminous and ultraluminous infrared galaxies (U/LIRGs) at unexpectedly high rates. To investigate this population of transients, we performed a search in mid-IR data from the Wide-field Infrared Survey Explorer (\textit{WISE}) satellite and its NEOWISE survey to detect and characterise luminous and smoothly evolving transients in a sample of 215 U/LIRGs. We report three new transients, all with $\Delta L > 10^{43}$~erg~s$^{-1}$, in addition to two previously known cases. Their host galaxies are all part of major galaxy mergers, and through radiative transfer model fitting we find that all have a significant contribution from an active galactic nucleus (AGN). We characterised the transients through measurements of their luminosities and resulting energetics, all of which are between 10$^{50.9}$~erg and 10$^{52.2}$~erg. The IR emission of the five transients was found to be consistent with re-radiation by the hot dust of emission at shorter wavelengths, presumably originating from an accretion event, onto the supermassive black hole. The corresponding transient rate of (1.6-4.6)$\times$10$^{-3}$  / yr / galaxy is over an order of magnitude higher than the rate of large amplitude flares shown by AGN in the optical. We suggest that the observed transients are part of a dust-obscured population of tidal disruption events (TDEs) that have remained out of the reach of optical surveys due to the obscuring dust. In one case, this is supported by our radio observations. We also discuss other plausible explanations. The observed rate of events is significantly higher than optical TDE rates, which can be expected in U/LIRG hosts undergoing a major galaxy merger with increased stellar densities in the nuclear regions. Continued searches for such transients and their multi-wavelength follow-up is required to constrain their rate and nature.}


   \maketitle
%

\section{\label{sec:intro}Introduction}

Tidal disruption event (TDE) candidates in dust-free nuclear environments of E+A/post-starburst galaxies are routinely discovered by optical wide-field surveys such as the All-Sky Automated Survey for Supernovae (ASAS-SN) and Zwicky Transient Facility (ZTF) \citep[see][and references therein]{Gezari2021}. However, TDE candidates in dusty environments can be much more challenging to identify and are often out of the reach of optical surveys. In particular, only three TDE candidates have been found in the nuclei of luminous and ultraluminous infrared galaxies (U/LIRGs), which are dust-obscured systems that host a powerful starburst and/or an active galactic nucleus (AGN), and often result from mergers of galaxies (see \citet{pereztorres2021} and references therein).

The only TDE candidate discovered so far in a U/LIRG at optical wavelengths was reported by \citet{tadhunter2017} in IRAS F01004-2237. The TDE interpretation of this object is based on strong and variable broad He lines and a TDE-like absolute optical magnitude, but the optical light curve of the transient was slower and longer lasting than canonically expected for a TDE. The optical event was followed by delayed, luminous, and slowly evolving infrared (IR) emission, with over 10$^{52}$~erg radiated \citep{dou2017}. \citet{tadhunter2017} suggested a TDE rate within U/LIRGs of 10$^{-2}$ TDE LIRG$^{-1}$ year$^{-1}$ based on this one event. However, its identification as a TDE is debated. Due to the slow spectral evolution of this and two other nuclear transients with similar properties, \citet{trakhtenbrot2019} argue against a TDE origin and suggest some other long-term intensified accretion events to explain their properties. More recently, this sub-class was expanded by \citet{Frederick2021} with more objects that showed Bowen fluorescence features and were associated with narrow-line Seyfert 1 AGN hosts. Most recently, \citet{cannizzaro2021a} and \citet{Tadhunter2021} both presented further spectroscopic observations of IRAS F01004-2237 and challenged the interpretation of \citet{trakhtenbrot2019}, instead supporting a TDE origin for the transient.

\citet{Mattila2018}, hereafter \citetalias{Mattila2018}, presented the discovery of an IR luminous transient (Arp 299-B~AT1) coincident with the nucleus B1 of the LIRG Arp 299, a result of systematic monitoring of a small sample of nearby LIRGs for dust-obscured supernovae (SNe) in the near-IR $K$ band \citep[see][]{Mattila2001}. Multi-wavelength follow-up included data in the near-IR, mid-IR, and radio, with non-detections at optical and X-ray wavelengths. The transient was extremely energetic, with more than 10$^{52}$~erg radiated in the IR over 10 years of evolution. Based on high-resolution radio very-long-baseline interferometry observations, \citetalias{Mattila2018} find Arp 299-B AT1 to be consistent with a TDE with a resolved and expanding radio jet. The viewing angle of the torus associated with the concealed AGN within this nucleus is constrained to be edge-on \citep{AlonsoHerrero2013}, and the inferred angle of the jet was not polar with respect to this torus, providing strong evidence that the transient was a TDE. The IR spectral energy distribution (SED) of the transient, and its evolution, was consistent with absorption and re-radiation of the TDE's UV/optical light by dust (an IR echo) in the polar regions of the dusty torus.

\citet{Kool2020}, hereafter \citetalias{Kool2020}, presented the discovery of another IR luminous nuclear transient at the centre of the northern nucleus in the LIRG IRAS 23436$+$5257, a result of adaptive optics assisted near-IR monitoring of $\sim$40 nearby LIRGs for dust-obscured SNe \citep{Kankare2008,Kankare2012,Kool2018}. AT 2017gbl was luminous in the IR and radio, but faint in the optical due to dust extinction. Given the combined IR and radio evolution, \citetalias{Kool2020} find a TDE to be the most plausible scenario, with an energetic SN ruled out based on the radio observations, and a changing-look active galactic nucleus (CLAGN) origin disfavoured based on the IR properties. They estimated the rate of AT 2017gbl-like events in LIRGs to be 10$^{n}$ TDE LIRG$^{-1}$ year$^{-1}$, where $n=-1.9^{+0.5}_{-0.8}$.

It is important to better constrain the rate of such optically obscured TDE candidates in LIRGs and ULIRGs, which are beyond the reach of any wide-field optical surveys. Based on a simulation of major mergers of galaxies, we could expect an increase in the TDE rate by two orders of magnitude over a period of 10 Myr compared to isolated galaxies \citep{Li2019}. Tidal disruption events provide an important means of feeding supermassive black holes (SMBHs) and can also play a role in driving outflows and regulating star formation if very common within LIRG nuclei. Therefore, constraining their rate in galaxy mergers and LIRGs has important implications for understanding galaxy evolution.

In this work we consider the mid-IR data available from the NEOWISE survey conducted by the Wide-field Infrared Survey Explorer (\textit{WISE}) satellite for a large sample of U/LIRGs and look for transients similar to those discussed above. In doing so, we discovered several new transient objects. We discuss the origin of individual objects and set a new rate for such transients within the U/LIRG population.

\section{\label{sec:SampObsDat}Galaxy data}%

\subsection{\label{subsec:sample}Luminous/ultraluminous infrared galaxy sample}

For our sample, we took all the U/LIRGs listed in the IRAS Revised Bright Galaxy Sample \citep[RBGS;][]{Sanders2003}. We updated the distances for the targets, assuming a $\Lambda$ cold dark matter cosmology with H$_0$=70.0 kms$^{-1}$Mpc$^{-1}$, ~$\Omega_{\Lambda}=0.721,$ and ~$\Omega_{M}$=0.279 \citep{Hinshaw2013} and taking distance values from the NASA/IPAC Extragalactic Database (NED), which uses the method described in \citet{Mould2000} to compensate for Virgo, Great Attractor, and Shapley cluster infall. We recalculated the L$_{IR}$ based on the updated distance and consider those galaxies with log$_{10}($L$_{IR}/L_{\odot})$ > 11 to be U/LIRGs, where L$_{IR}$ is the $8-1000~\mu$m luminosity. We removed 2 objects that were mis-categorised as galaxies in the RBGS (IRAS F05170+0535 is the T Tauri star HD 34700 \citep{Sterzik2005} and IRAS 05223+1908 is a young stellar object \citep{Chu2017}). The resulting sample is listed in Table \ref{tab:sample} and consists of 215 galaxies, after excluding NGC 1068 (see Sect. 2.2). We use the names given by the RBGS except in the case of NGC 3690/IC 694, which we refer to as Arp 299, in order to make clear the association with the transient Arp 299-B~AT1.

\subsection{\label{subsec:WISE}WISE data}
Repeated mid-IR measurements of our sample galaxies come from data taken by the \textit{WISE} satellite. \textit{WISE} initially surveyed the sky at 3.4, 4.6, 12, and 22 $\mu$m (channels $W1$ - $W4$) in 2010. After depletion of its cryogen, the survey was continued through the post-cryogenic and NEOWISE surveys with channels $W1$ and $W2$ only \citep{Mainzer2011,Mainzer2014}. We took all the available data from the public AllWISE Multiepoch Photometry Table and the NEOWISE-R Single Exposure (L1b) Source Table\footnote{\url{https://irsa.ipac.caltech.edu/cgi-bin/Gator/nph-scan?mission=irsa&submit=Select&projshort=WISE}} for galaxies in our sample, which provide observations of all our targets at a 6-month observing cadence, with an approximately 3 year gap in observations from 2010-2013, between the initial and subsequent missions. The NEOWISE-R data consist of all data releases available at the time of writing, consisting of observations taken from 13 December 2013 and 13 December 2021.

Many U/LIRGs are part of interacting systems and may have multiple nuclei that can host transient events \citep{pereztorres2021} similar to those reported by \citetalias{Mattila2018} and \citetalias{Kool2020}. In order to identify galaxies with multiple nuclei detected with WISE, we considered all the objects given in the AllWISE \citep{Wright2010,Mainzer2011} catalogue with $W1$ < 15 mag that were within 60" of the coordinates for the U/LIRG given by the RBGS catalogue. The magnitude cutoff was selected as the approximate magnitude where the signal-to-noise ratio (S/N) = 10 for PSF photometry with NEOWISE, in order to remove spurious detections. In practice, none of the LIRG nuclei in our sample have mean $W1$ magnitude > 14 mag. The resolution of the IRAS satellite was 0.5$\arcmin$ at best and as much as 2$\arcmin$ at longer wavelengths, whereas the resolution of the WISE satellite is 6.1$\arcsec$ and 6.4$\arcsec$ at 3.4 $\mu$m and 4.6 $\mu$m, respectively. We therefore allowed a large matching radius to include separate nuclei that are resolvable with WISE but may have been unresolved in the RBGS catalogue. We then queried these sources in the NED to determine if they were identified as additional nuclei associated with the same galaxy, and inspected images of the galaxies to confirm these identifications. There were three sources where a companion galaxy was clearly visible and was detected with NEOWISE, but was not present in the AllWISE catalogue -- these are included in the sample. There were 55 galaxies for which we identified multiple nuclei through this process and they are noted in Table \ref{tab:sample} with the AllWISE name (where available) and coordinates used for each source given.

As described above, the $W1$ and $W2$ NEOWISE data were taken from the online WISE catalogue service. The WISE survey strategy results in approximately a dozen repeated observations of an object at a daily cadence during each bi-yearly epoch of observation. We adopted the median value of the individual measurements at each visit as the magnitude, in order to be robust to outliers and increase the S/N of the observation. All magnitudes in this work are given in the Vega system. We removed any individual measurements that were further than 2$\arcsec$ from the identified source coordinates. There are a number of quality flags provided with the WISE data, which indicate various sources of inaccuracy in the measurements. For the majority of targets, we excluded any measurements that were flagged this way. However, we found that in cases where nearly all available data on a target were given the `contamination and confusion flag', the measurements nevertheless appeared accurate, and we decided to make use of the data. This is most notably the case for Arp 299-B AT1 and these data are discussed in Sect. \ref{subsec:Arp299}.

For the uncertainty, we adopted the standard error of mean of the individual measurements taken at each bi-yearly epoch of WISE observation after three sigma-clipping outliers. We added 0.0026 mag and 0.0061 mag uncertainties in quadrature to the $W1$ and $W2$ measurements, respectively, which are the RMS residuals found in the photometric calibration during the survey period\footnote{\url{https://wise2.ipac.caltech.edu/docs/release/neowise/expsup/sec4_2d.html}}. For $W2$, there is a seasonal variation of about 0.02 mag in this calibration due to heating of the satellite during the equinox, which we do not correct for, but this is insufficiently large to affect our results.

In cases where the M$_{W1}$ < 8 mag or M$_{W2}$ is < 7 mag, the NEOWISE source catalogue overestimates the magnitude of the measured source, and this effect increases with greater source brightness. In these cases we applied a saturation correction\footnote{The saturation correction is described here: \url{https://wise2.ipac.caltech.edu/docs/release/neowise/expsup/sec2_1civa.html}}, and the uncertainty from this was added in quadrature to the other uncertainties. This correction is required for four objects: Arp 299, NGC 1068, UGC 08058, and NGC 1365. However, NGC 1068 is significantly brighter than the saturation limit, introducing very large uncertainties, and therefore we excluded it from the sample.

\subsection{\label{subsec:filtering}Transient detection}

\begin{table}
\caption{Galaxies with $\Delta L > 10^{43}$ erg s$^{-1}$ in either WISE band.}
\centering
\resizebox{\hsize}{!}{
\begin{tabular}{lcccc} \hline \hline
Galaxy &  $\Delta L_{W1}$ & $\Delta L_{W2}$ & Host Classification \\
 & log$_{10}$(erg~s$^{-1}$) & log$_{10}$(erg~s$^{-1}$)   \\ \hline
AM 0702-601N & 43.60 & 43.55 & S2  \\ 
Arp 299 & 43.18 & 43.28 & H2  \\ 
ESO 286-IG019 & 42.53 & 43.18 & H2   \\ 
IRAS F05189$-$2524  & 43.85 & 43.74 & S1h \\ 
IRAS 23436$+$5257  & 43.20 & 43.09 & -- &  \\ \hline
NGC 1275 & 43.36 & 43.30 & S1.5  \\ 
NGC 7469 & 43.59 & 43.51 & S1.5 \\ 
NGC 7674E & 43.06 & 42.92 & S1h  \\ 
UGC 05101 & 43.05 & 43.38 & S1  \\ 
UGC 08058 & 44.39 & 44.19 & S1 \\ \hline
\end{tabular}
}
\tablefoot{Host spectroscopic classification is taken from \citet{Veron2010}, where available. S1 and S2 refer to Seyfert 1 and 2, respectively; S1.5 are intermediate objects; S1h are S2 objects with broad lines visible in polarised light; and H2 refers to objects with a nuclear H II region. Galaxies below the horizontal line show AGN-like variability in WISE data and are excluded from further analysis.}
\label{tab:candidates}
\end{table}

\begin{figure}
\centering
   \resizebox{\hsize}{!}{\includegraphics[width=0.5\textwidth]{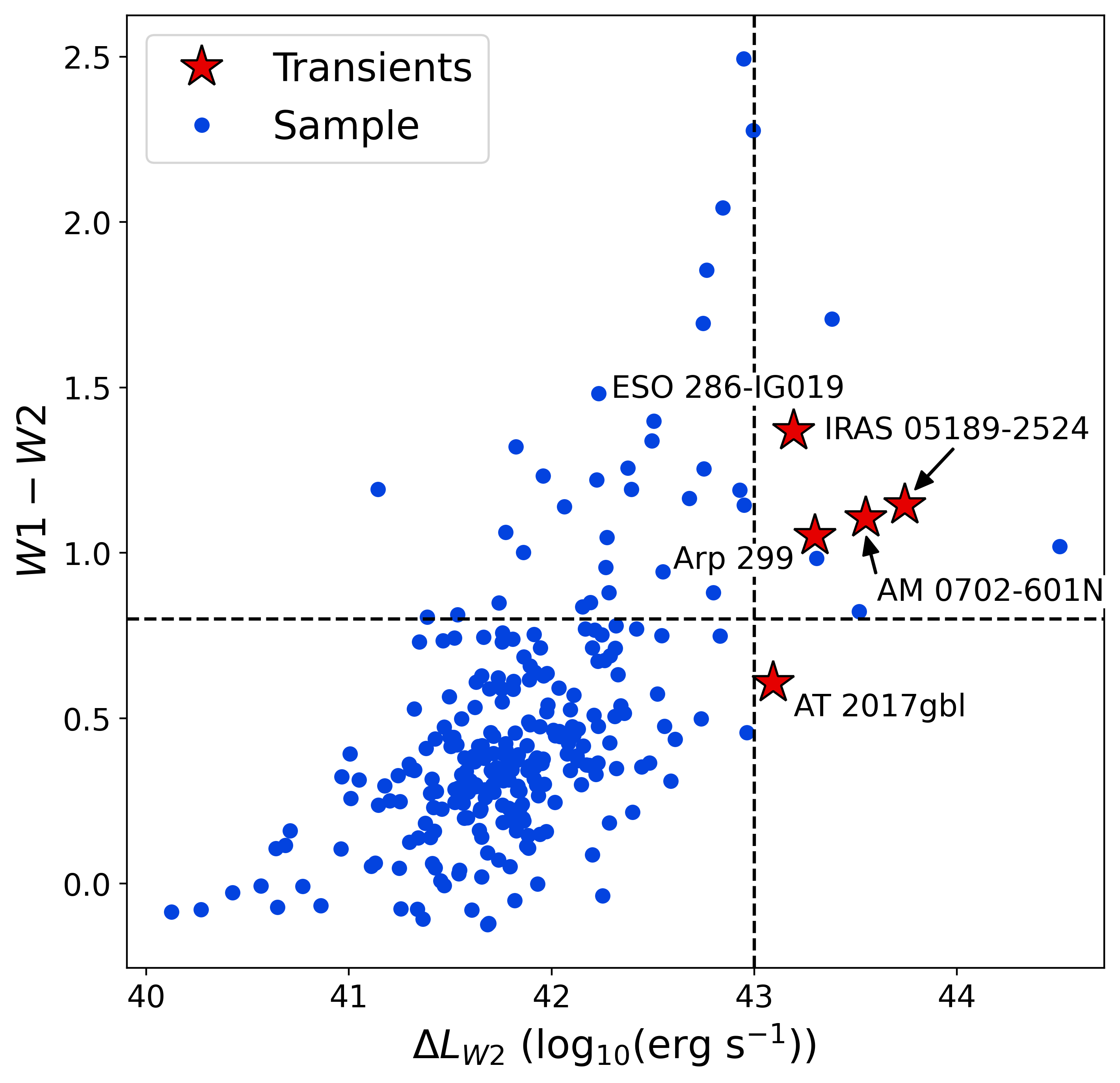}}
     \caption{$W1-W2$ colour of the galaxies in the sample, plotted against $\Delta L_{W2}$. The vertical dashed line shows the luminosity criterion that we used to detect transients in our sample. A commonly used criterion to detect the presence of an AGN is if $W1-W2$ > 0.8 \citep{Stern2012}, shown here by the horizontal line. For the five transients, the colour shown is derived from the adopted host contribution, as described in Sect. \ref{detections}.}
     \label{fig:WISE_colours}
\end{figure}   

Within our sample, we carried out a search to find transients similar to the three TDE candidates discovered in U/LIRGs so far: Arp 299-B AT1, AT 2017gbl, and the transient in IRAS F01004-2237. These events are all very luminous, differentiating them from SNe, and consist of a single major outburst that evolves smoothly, rather than the stochastic variability that characterises AGN \citep{koshida2014}. We apply a luminosity constraint and measure the difference between the least and most luminous observation within the NEOWISE data: $\Delta L = L_{\text{max}} - L_{\text{min}}$. We calculate $\Delta L$ from the NEOWISE data only, to avoid our results being affected by potential systematic offsets in the measurements between the different WISE surveys. We then consider objects with $\Delta L > 10^{43}$~erg~s$^{-1}$ in either the WISE $W1$ or $W2$ band as potential transients. The least luminous event in NEOWISE of the previously mentioned TDE candidates was AT 2017gbl, with $\Delta L_{W2} = 10^{43.1}$~erg~s$^{-1}$. This cut will also exclude bright SNe, as the most luminous SN observed in the mid-IR as a part of a large sample of Spitzer observed SNe, SN 2010jl, has $L_{3.6\mu \text{m}} \text{ and } L_{4.5\mu \text{m}} < 10^{42.3}$~erg~s$^{-1}$ \citep{Fransson2014,Szalai2019}.

We find ten U/LIRGs in our sample that have $\Delta L > 10^{43}$~erg~s$^{-1}$in the $W1$ or $W2$ band. These objects are listed in Table \ref{tab:candidates}. The objects in the lower section of the table are ruled out as being single outbursts, as they all fluctuate around their mean luminosity, rather than showing a single smoothly evolving event. This behaviour is consistent with AGN variability, and all these targets have hosts that are classified as Seyfert 1 galaxies in the catalogue of \citet{Veron2010}. The light curves of these objects are shown in Fig. \ref{fig:Not Transients}. The candidate UGC 08058 is rejected based on the very large scatter in the individual WISE measurements and corresponding very large uncertainty of $\sim$0.1 mag in the measurements. This object also has a Seyfert 1 host galaxy (also known as Mrk 231). The remaining objects all show a clear single outburst, but otherwise have varied properties, which we discuss below. They include the aforementioned Arp 299-B AT1 and AT 2017gbl in IRAS 23436$+$5257, which are both recovered by our detection criteria.

\begin{figure*}
\centering
   \includegraphics[width=\textwidth]{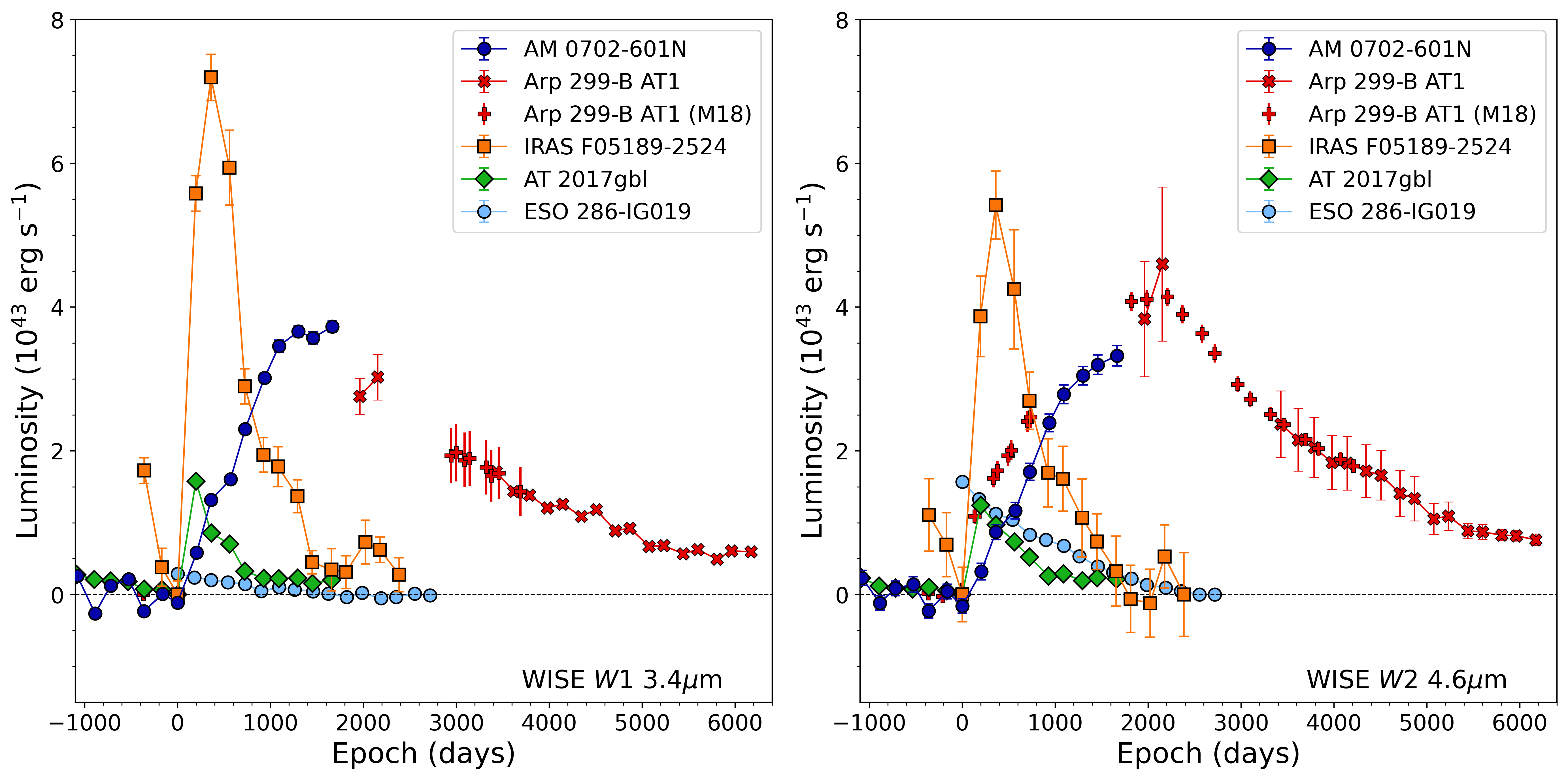}
     \caption{Luminosity evolution of the detected transients in WISE $W1$ and $W2$ bands. Additionally included are host-subtracted 3.6 and 4.5 $\mu$m measurements of Arp 299-B AT1 taken with Spitzer, from \citetalias{Mattila2018}. The host contribution for subtraction and the zero epochs adopted are described in Sect. \ref{detections}. }
     \label{fig:Transients}
\end{figure*}   

Figure \ref{fig:WISE_colours} shows the mean NEOWISE $W1-W2$ colours of our sample galaxies compared to their $\Delta L_{W2}$ values. For the galaxies with a single outburst, we use the host magnitudes without the transient for the colour calculation (The values adopted are described in Sect. \ref{detections}). The host galaxies of four of the five transients we identify have $W1-W2$ > 0.8, which is a criterion commonly used to identify hidden AGN \citep{Stern2012}. We investigate the nature of the host galaxies further in Sect. \ref{subsec:SED}.

\section{\label{detections}Transients}

In this section we describe each of the transients and their host galaxies in turn, followed by analysis of their properties. In order to consider the properties of the transients without the contribution from the host galaxy, we adopt the following host contributions for subtraction throughout this work: the final measurement before the outburst in the case of IRAS~F05189$-$2524 and AT 2017gbl; the mean of the pre-outburst points for AM~0702-601N; and the mean of the final two observations for ESO~286-IG019. The $W1$ observations of Arp~299-B~AT1 assume the same host contribution as derived by \citetalias{Mattila2018} from the Spitzer observations, while a correction is applied to the $W2$ data based on matching the WISE and Spitzer measurements where they overlap (see Sect. \ref{subsec:Arp299}). The zero epochs adopted are the last pre-outburst observation for AM 0702-601N, IRAS F05189$-$2524 and AT 2017gbl; the first NEOWISE measurement for ESO~286-IG019; and the discovery epoch found in \citetalias{Mattila2018} for Arp~299-B~AT1. 

The light curves for the transients, with the assumptions described above, are shown in Fig.~\ref{fig:Transients} and a summary of their key observational parameters can be found in Table~\ref{tab:Parameters}. We also include these parameters for some transients from the literature in Table~\ref{tab:Parameters}. These transients are of interest for comparison based on either their physical mechanism or host galaxy, and will be discussed in Sect. \ref{subsec:transientInterp}.

\subsection{\label{AM0702}Transient in AM 0702-601N}

\begin{table*}
\caption{Parameters derived for the transients in our sample, as well as for other transients taken from the literature.}
\centering
\begin{tabular}{l c c c c c c c} \hline \hline
Transient & Peak L$_{3.4\mu m}$ & Peak L$_{4.6\mu m}$  & Zero epoch & Rise time  & Total Energy  & Max BB T & Ref \\ 
 & log$_{10}$(erg~s$^{-1}$) &   log$_{10}$(erg~s$^{-1}$) & MJD & days &  log$_{10}$(erg) & K &  \\ \hline

IRAS F05189$-$2524 & 43.9 & 43.7 & 57081 & 360 & 51.8 & 1470$^{+400}_{-220}$  &  (1) \\ 
AM 0702-601N & >43.6 & >43.5 & 57870 & >1665 & >51.7 & 1590$^{+380}_{-230}$   & (1) \\ 
ESO 286-IG019 & 42.5 & 43.2 & 56776 & -- & -- & --   & (1) \\
AT~2017gbl & 43.2  & 43.1  & 57754 & 197 & 50.9 & 1200$^{+40}_{-30}$   & (1) \\ 
Arp 299-B AT1 & 43.5 & 43.7 & 53361 & 2208$^*$ & >52.2 & 1045$^{+7*}_{-7}$  & (1) , (2) \\
\cline{1-8}
IRAS F01004-2237 & 44.2 & 44.3 & -- & >2183 & >52.1 & 850  & (3) \\ 
SN~2010jl & 42.3 & 42.2 & -- & $\sim$600 & 50.43 & 2040  & (4) \\ 
ASASSN-14li & 41.3 & 41.1 & -- & <21 & 49.5 & 1340$^{+276}_{-276}$  & (5) \\ 
ASASSN-15lh & 43.2 & 43.2 & -- & 562 & 51.4 & 1360$^{+330}_{-330}$ & (5) \\ 
\hline
\end{tabular}
\tablefoot{Quiescent values for host subtractions are as described in Sect. \ref{detections}. The measurement uncertainties for the peak luminosity, derived purely from the uncertainty of the peak flux measurement, are of order 10$^{-2}$ log$_{10}$(erg~s$^{-1}$) in all but one case. However, the total uncertainty is dominated by the systematic uncertainty in the choice of the quiescent epoch. The time to peak is the difference between the observed brightest point and the listed zero epoch, in cases where the peak is visible. The total energy and maximum blackbody (BB) temperature are derived from BB fitting to the data, further described in Sect. \ref{subsec:BB_fitting}. The total energy assumes a linear change in luminosity between the epochs.}
\tablebib{(1)  This work; (2) \citet{Mattila2018} (denoted by an asterisk); (3) \citet{dou2017}; (4) \citet{Fransson2014}; (5) \citet{Jiang2021b}.}
\label{tab:Parameters}
\end{table*}

AM 0702-601N has coordinates RA=$07^{\textrm{h}} 04^{\textrm{m}} 05^{\textrm{s}}.250$, Dec= $-60^{\circ} 19\arcmin 58\arcsec.43$ and is located at a distance of 137.1 $\pm$ 9.7 Mpc. Assuming this distance, the galaxy has log$_{10}$(L$_{IR}/L_{\odot}$)= 11.64. The galaxy has a distant companion, known as AM~0702-601S, which is separated 1.5$\arcmin$ to the south, corresponding to a projected distance of $\sim$60~kpc. AM~0702-601N shows signs of interaction with this companion, through expanded outer spiral arms and a tidal tail. These two galaxies were not resolved at longer wavelengths with the resolution of the IRAS satellite, but the northern galaxy, where the transient is located, is brighter in all IR bands. \citet{Haan2011} find a black hole (BH) mass of log$_{10}$($M_{BH}/M_{\odot}$) = 8.09$\pm$0.11 for this galaxy, based on the bulge luminosity and using the Marconi-Hunt relation \citep{Marconi2003}.

The WISE light curves of AM 0702-601N show minor variation up until MJD = 57870d (27 April 2017). The next observation shows a larger increase than the previous variation in both the $W1$ and $W2$ bands, and the luminosity then continues to smoothly increase for a total of $\sim$4.5 years across nine WISE epochs. The latest three epochs show a decline in the rate of luminosity increase, particularly in $W1$, presumably as the transient approaches the peak. The rise of the light curve is quite similar to the luminous and long lived transient Arp 299-B AT1 \citepalias{Mattila2018}, although the transient in AM 0702-601N has a slightly bluer colour. 

We obtained optical observations in the $V$ band of the galaxy from the Catalina Real-time Transient Survey \citep[CRTS;][]{Drake2009}, covering the period from 5 October 2005 to 3 May 2013, or $-$4221 d to $-$1455 d compared to the adopted explosion date for the transient. There is no indication of significant variability during this period, given the photometric uncertainties. These observations do not overlap with the transient, but indicate that the level of historical optical variability in the galaxy was low. This can indicate either a real lack of variability or that a large amount of dust extinction concealed any optical variability.

\subsection{\label{subsec:IRAS05189}Transient in IRAS F05189$-$2524}

IRAS F05189$-$2524 has coordinates RA=$05^{\textrm{h}} 23^{\textrm{m}} 03^{\textrm{s}}.936$, Dec=$-25^{\circ} 19\arcmin 00\arcsec.08$ and is located at a distance of 180.9 $\pm$ 12.7 Mpc. Assuming this distance, the galaxy has log(L$_{IR}$/L$_{\odot}$)= 12.17 and is therefore classified as a ULIRG. The morphology of the galaxy, showing a compact core along with two curved tail structures, indicates an advanced major merger product \citep{Sanders1988,Veilleux2002}.

\citet{Xu2017} report a SMBH mass for this galaxy of log$_{10}$($M_{BH}/M_{\odot})$= 8.62, using central velocity dispersion measured from the Ca {\sc ii} triplet line widths \citep{Rothberg2013} and the $M_{BH} - \sigma$ relation \citep{Tremaine2002}. However, \citet{Dasyra2006} find a much lower value of log$_{10}$($M_{BH}/M_{\odot})$= 7.47, derived from their velocity dispersion measured from CO band-head features in the near-IR. This is related to the `$\sigma$ discrepancy', discussed in \citet{Rothberg2013}, in which the spectral measurements are sampling stellar populations with different ages. As the Ca {\sc ii} triplet measurements sample an older stellar population than the CO band-heads, it is more likely that the larger $M_{BH}$ estimate is accurate. There is evidence from the X-ray spectrum of the galaxy that the SMBH has a high spin, with some of the models presented in \citet{Xu2017} finding a dimensional spin parameter $>$0.9, although others favour a lower value.

The WISE light curves for this galaxy show that after a period of decline, a bright transient occurred between MJD 57080.6 and 57275.9 (27 February 2015 to 10 September 2015). The transient peaked on MJD 57441.0, implying a rise time of 0.5-1 year. The transient declines relatively rapidly for a year, and then more slowly. Recent observations 
show flux densities close to the level observed immediately before the outburst, although with some small fluctuations, implying the galaxy nucleus may have returned to a `quiescent' level after 4.5 years.

This galaxy was observed as part of the VISTA Hemisphere Survey in the near-IR $J$ and $K$ bands. By a fortunate coincidence, these three observations from 29 December 2015, 30 December 2015, and 28 January 2016 took place close to the measurements of the transient's peak in the NEOWISE light curve (23 February 2016). We stacked the two observations from December 2015 to obtain the deepest possible image. We observed the host with the Nordic Optical Telescope (NOT)/NOTCam as part of the NOT Unbiased Transient Survey 2 (NUTS2) programme on 2 November 2020. This is 1714 days after the transient's peak, and the host nucleus is quiescent in the WISE data at this point, so we assume the transient has already faded. We then measured the galaxy flux in the transient images with aperture photometry. We performed an arithmetic subtraction on the resulting flux measurements of the galaxy and recovered a $J-$band magnitude for the transient of m$_J$ = 14.50$\pm$0.02, which we include in our blackbody (BB) fits in Sect. \ref{subsec:BB_fitting}. Unfortunately the VISTA $K$ band data were saturated and unusable.

We checked optical observations in the $V$ band of the galaxy from the CRTS. 
As in the case above for AM~0702-601N, these observations do not overlap with the transient event, but they indicate that historical optical variability in the galaxy was insignificant.


\subsection{\label{subsec:ESO286}Transient in ESO 286-IG019}

ESO 286-IG019 (also known as IRAS 20551-4250) has coordinates RA=$20^{\textrm{h}} 58^{\textrm{m}} 27^{\textrm{s}}.818$, Dec=$-42^{\circ} 38\arcmin 59\arcsec.42$ and is located at a distance of 187 $\pm$ 13 Mpc. Assuming this distance, the galaxy has log(L$_{IR}$/L$_{\odot}$)= 12.07 and is also classified as a ULIRG. The galaxy displays a compact core and tidal tails, one of which is very prominent \citep{Johansson1991} and is classified as a post merger by \citet{Haan2011}.

\citet{Haan2011} find a BH mass of log$_{10}$($M_{BH}/M_{\odot}$) = 8.64 $\pm$ 0.08 for this galaxy, based on the bulge luminosity and using the Marconi-Hunt relation \citep{Marconi2003}. \citet{Dasyra2006} find a much lower value of log$_{10}$($M_{BH}/M_{\odot}$) = 7.51 derived from velocity dispersion measured from CO band-head features in the near-IR and the $M_{BH} - \sigma$ relation from \citet{Tremaine2002}. There is again evidence of the $\sigma$ discrepancy, with a larger velocity dispersion derived from the Ca {\sc ii} lines reported by \citet{Rothberg2010}, which would correspond to log$_{10}$($M_{BH}/M_{\odot}$) = 8.19.

The NEOWISE light curves for this galaxy show a steady decline in both bands from the first observations until the most recent pair of observations in 2021, in which the measurements in both bands have levelled out. The galaxy only passes our transient detection limit in the $W2$ band, while $\Delta L_{W1}=10^{42.5}$ erg s$^{-1}$, which is significantly less than required to pass the cut. If we take the mean of the final two observations to represent the quiescent host galaxy flux, the transient is much redder than the other transients in the sample. Although there appears to be a levelling out of the decline in the most recent data, we cannot be certain that our last data represents the quiescent host galaxy flux, so it is difficult to determine if the colour is physical or due to an incorrect subtraction of the host flux.

\subsection{\label{subsec:Arp299}Arp 299-B AT1}
Arp 299-B AT1 was discovered in the near-IR $K$ band in 2005 and was followed up at mid-IR wavelengths by Spitzer until 2016. These data, along with observations at other wavelengths, are discussed in depth in \citetalias{Mattila2018}. The transient has also been observed by NEOWISE and the $W1$ points agree remarkably well with the Spitzer 3.6 $\mu$m light curve. As there are no WISE observations of Arp 299 in quiescence, we make use of the Spitzer quiescent host values given in \citetalias{Mattila2018} for host subtraction. As the $W1$ measurements agree with the Spitzer values, no correction was required. The $W2$ points are offset compared to the Spitzer 4.5 $\mu$m data. The discrepancy could be due to the non-linearity of the $W2$ NEOWISE data, differences between the Spitzer and WISE filters, and/or differences in the spatial resolution. To obtain a quiescent host value for the $W2$ data, we calculated the mean difference between the Spitzer 4.5 $\mu$m and $W2$ observations in the region where the two datasets overlap, where there are five NEOWISE measurements. The resulting correction of 0.0438 $\pm$ 0.0023 Jy was applied to the Spitzer 4.5 $\mu$m quiescent host value.

The NEOWISE data show the mid-IR light curve decline continuing at a similar rate as shown by the pre-2016 light curve. Additionally, there are two epochs of observation from the original WISE mission at the peak of the transient. The $W2$ measurements agree well with the Spitzer values after the correction was applied, although as these data required a non-linearity correction, they have large uncertainties.

\subsection{\label{subsec:17gbl}AT~2017gbl}

The transient AT~2017gbl in the LIRG IRAS 23436+5257 is described in detail in \citetalias{Kool2020}. Here we report two additional epochs of WISE imaging released since the publication of that work. These measurements were analysed following the same procedures as reported in \citetalias{Kool2020}. The light curves seem to have levelled out, but to a slightly higher level than the quiescent flux implied by the last observation before the rise of the transient.

\subsection{\label{subsec:Radio}Radio observations}

\begin{figure}
   \centering
   \resizebox{\hsize}{!}{\includegraphics{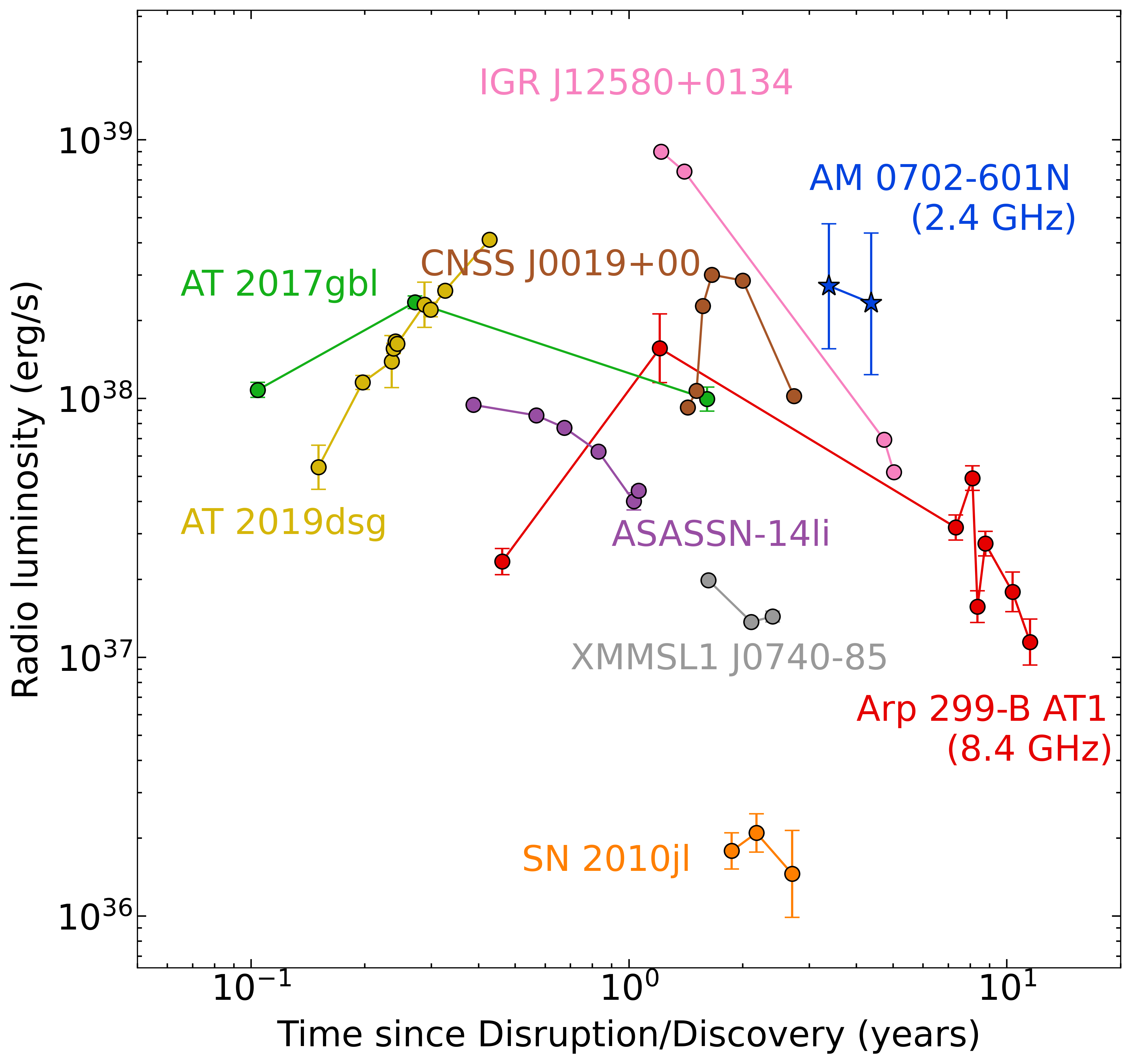}}
      \caption{Radio luminosity of AM 0702-601N compared with a sample of literature TDEs as well as with SN~2010jl and AT~2017gbl. Observations are at $\sim$5 GHz, except where noted in the plot. The style of the plot is taken from Fig. 1 of \citet{Alexander2020}. The data sources are: \citet{cannizzaro2021b} for AT~2019dsg; \citet{Irwin2015} and \citet{Perlman2017} for IGR~J12580+0134; \citet{Anderson2020} for CNSS~J0019+00; \citet{Alexander2016} for ASASSN-14li; \citet{Alexander2017} for XMMSL1~J0740-85; \cite{Fransson2014} for SN~2010jl; \citetalias{Mattila2018} for Arp~299-B AT1; \citetalias{Kool2020} for AT 2017gbl; and this work for AM 0702-601N.}
         \label{fig:radio}
  \end{figure}

In the case of AM 0702-601N, we obtained radio observations of the host galaxy using the Australia Telescope Compact Array (ATCA) with 2048$\times$1~MHz bandwidths centred on 5.5 and 9.0 GHz, and 512$\times$1~MHz bandwidths centred at 1.3, 1.8, 2.4, and 2.9 GHz, in two epochs of observation a year apart (11 September 2020 and 11 September 2021). During these runs the ATCA primary flux calibrator PKS B1934-638 was observed at each frequency for absolute and bandpass calibration, while the nearby source J0625--6020 was observed regularly for gain and phase calibration. The ATCA data were processed using tasks within the {\sc miriad} package \citep{Sault1995}. After editing and calibration of the data, images at each frequency were made using multi-frequency synthesis and a robust weighting factor of 0.5, then cleaned to a level approaching the r.m.s. noise. Fitting a Gaussian point source at the location of AM 0702-601N yielded the flux densities listed in Table~\ref{tab:radio_AM0702}, with the uncertainties given by the quadrature sum of the image r.m.s. and a fractional error on the absolute flux scale in each band \citep{Weiler2011}. Significant radio frequency interference on the mainly short baselines on 11 September 2020 precluded any useful measurement at 1.3 GHz.

We also searched for archival radio observations of the host galaxies of our newly discovered transients. In the case of AM 0702-601N, previous ATCA observations at 2.1 GHz with 32$\times$64~MHz bandwidths were available from 16 June 2012. These observations were processed following the procedures described above but using J0742-56 for gain and phase calibration. Radio emission was also detected at the position of AM~0702-601N in the Sydney University Molonglo Sky Survey (SUMSS; \citealt{SUMSS}), and in two overlapping fields of the Rapid ASKAP Continuum Survey (RACS; \citealt{RACS}).

No useful radio observations were available from the archives for the other sources and therefore we concentrate only on AM 0702-601N here. No significant variability was apparent at $\sim$0.9 GHz between a SUMSS observation before the outburst at $-$5354 days and RACS observations after the outburst at +738 days. However, comparing the ATCA measurement at 2.1 GHz taken before the outburst at $-$1776 days with our ATCA observations at 1.8 and 2.4 GHz from 2020/21 we see that the source has brightened by some 40\% in that time. Furthermore, we detected a statistically significant drop in the radio flux densities at 5.5 and 9.0 GHz between 1233 and 1597 days. At other frequencies the values of the flux densities were consistent with no variability between these two epochs.

In Fig. \ref{fig:radio} we show the host-subtracted radio luminosity of the transient in AM 0702-601N at $\sim$2.4 GHz. We compare this with the luminosities of a sample of radio detected TDEs (taken from \citet{Alexander2020}) and the comparison objects from Table~\ref{tab:Parameters} with radio detections. The transient in AM 0702-601N is comparable in luminosity to those TDEs whose radio emission is proposed to be driven by off-axis, initially relativistic jets, although it appears to be slightly more luminous. It is much less luminous than TDEs with radio emission proposed to arise from on-axis relativistic jets (for example, Sw J1644+57 \citep{Zauderer2011,Bloom2011} and Sw J1112$-$82~\citep{Brown2017}), which have typical radio luminosities of $\sim$10$^{40-41}$ erg s$^{-1}$. Future observations with ATCA should eventually allow the assessment and removal of the quiescent radio fluxes of the nucleus to analyse the radio flux densities at several frequencies free from the host contamination.

\begin{table}
\caption{Radio observations of AM 0702-601N.}
\centering
\resizebox{\hsize}{!}{
\begin{tabular}{lccccc} \hline\hline 
UT Date & MJD & Epoch & Array/Survey & Frequency & Total flux density \\ 
&&(days)&&(GHz)&(mJy) \\ 
(1)&(2)&(3)&(4)&(5)&(6) \\ \hline
\hline
30 Oct. 2002 & 52516 & $-$5354 & SUMSS & 0.843 & 48.2 $\pm$ 1.8 \\
16 June 2012 & 56094 & $-$1776 & ATCA 6D & 2.11 & 16.1 $\pm$ 1.6 \\
6 May 2019 & 58609 & +739 & RACS & 0.888 & 45.6 $\pm$ 0.2 \\
6 May 2019 & 58609 & +739 & RACS & 0.888 & 48.0 $\pm$ 0.2 \\
11 Sept. 2020 & 59103 & +1233 & ATCA 750B & 1.8 & 24.3 $\pm$ 2.7 \\
11 Sept. 2020 & 59103 & +1233 & ATCA 750B & 2.4 & 21.0 $\pm$ 2.1 \\
11 Sept. 2020 & 59103 & +1233 & ATCA 750B & 2.9 & 19.5 $\pm$ 2.0 \\
11 Sept. 2020 & 59103 & +1233 & ATCA 750B & 5.5 & 14.1 $\pm$ 0.9 \\
11 Sept. 2020 & 59103 & +1233 & ATCA 750B & 9.0 & 9.7 $\pm$ 0.5 \\
11 Sept. 2021 & 59468 & +1597 & ATCA 6A & 1.3 & 22.9 $\pm$ 2.3 \\
11 Sept. 2021 & 59468 & +1597 & ATCA 6A & 1.8 & 22.4 $\pm$ 2.3 \\
11 Sept. 2021 & 59468 & +1597 & ATCA 6A & 2.4 & 20.3 $\pm$ 2.0 \\
11 Sept. 2021 & 59468 & +1597 & ATCA 6A & 2.9 & 17.7 $\pm$ 1.8 \\
11 Sept. 2021 & 59468 & +1597 & ATCA 6A & 5.5 & 11.5 $\pm$ 0.6 \\
11 Sept. 2021 & 59468 & +1597 & ATCA 6A & 9.0 & 6.9 $\pm$ 0.4 \\
\hline

\end{tabular}
}
\tablefoot{Columns (1-3) list the observation date in UTC and MJD formats, and the time after the discovery; (4) the telescope and array configuration, or survey; (5) the central frequency; (6) the measured total flux density.}
\label{tab:radio_AM0702}
\end{table}


\subsection{\label{subsec:BB_fitting}Blackbody fitting for the transients}

\begin{figure*}
   \centering
   \includegraphics[width=\textwidth]{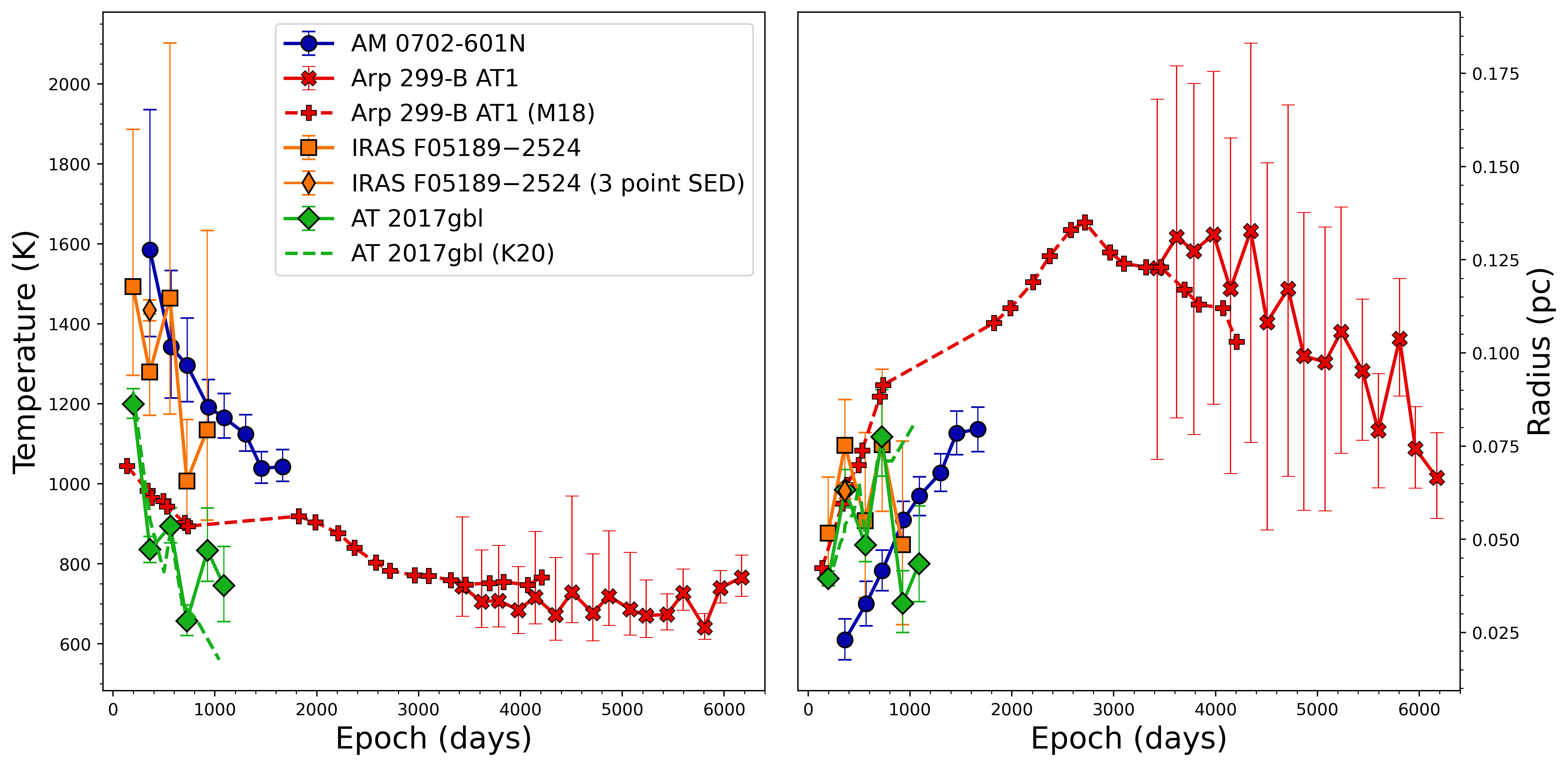}
      \caption{Inferred BB parameters (temperature and radius) for all the transients in the sample. The dashed lines show results taken from \citetalias{Mattila2018} and \citetalias{Kool2020}.}
         \label{fig:BB_results}
  \end{figure*}

\begin{figure*}
   \centering
   \includegraphics[width=\textwidth]{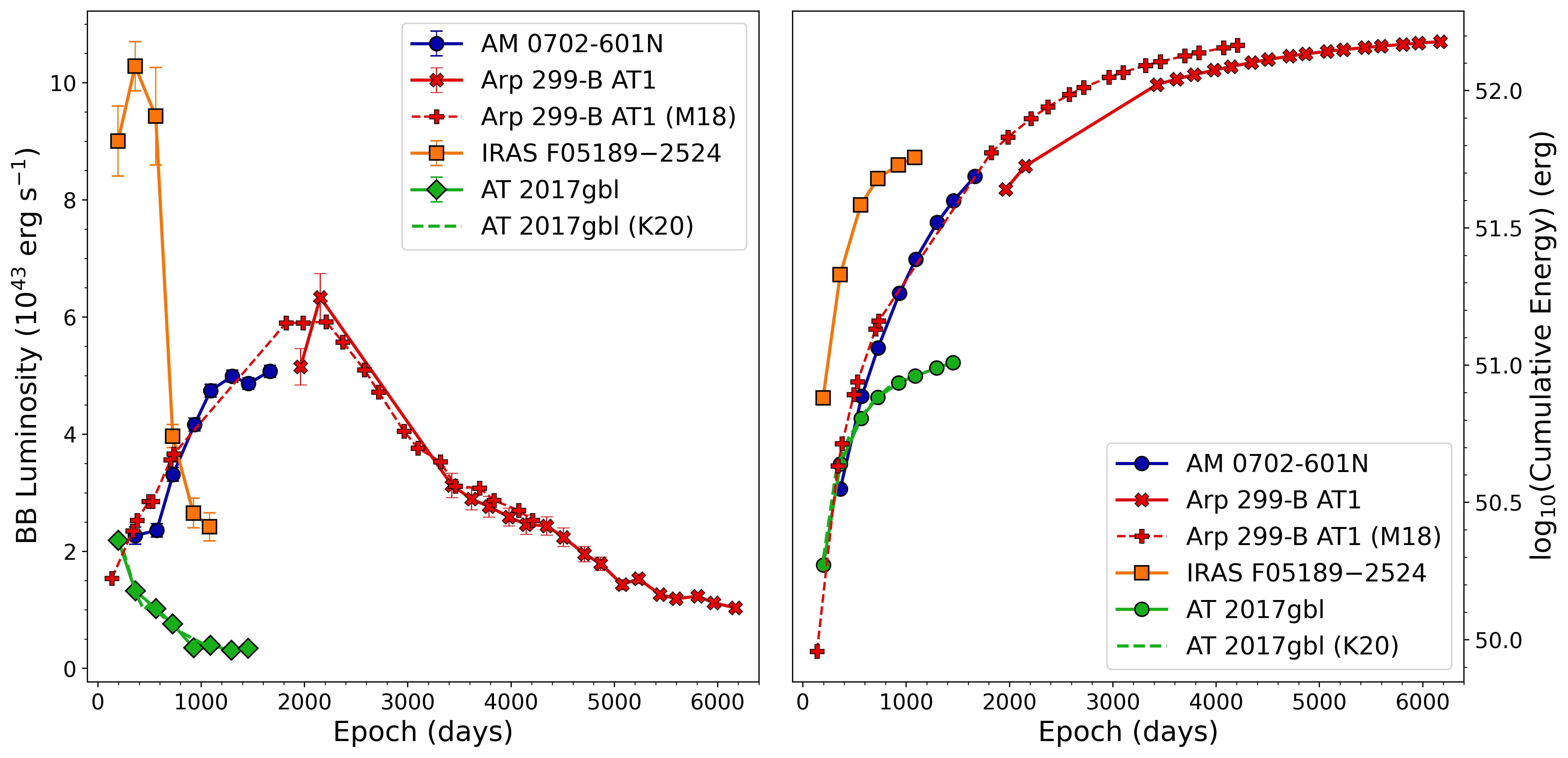}
      \caption{Energetics for the transients inferred from the BB parameters. Left panel: Luminosities inferred from BB parameters for all the transients in the sample. Right panel: Cumulative radiated energies inferred from the luminosities of the BB associated with the transients. The dashed lines show results taken from \citetalias{Mattila2018} and \citetalias{Kool2020} in both panels.}
\label{fig:BB_LE}
\end{figure*}

To evaluate the nature of these transients, we performed BB fitting. The IR outbursts detected by WISE are likely to be produced by IR echoes, in which dust is heated by the UV and optical emission from an energetic accretion event and re-radiates this energy in the IR. We expect that this radiation can be well described by a BB spectrum, and this has been observed in the case of the previously discovered transients in our sample, Arp 299-B AT1 and AT 2017gbl \citepalias{Mattila2018,Kool2020}, where good fits were obtained for the observed near-IR and mid-IR emission by a single component BB. In the case of the transients newly discovered here, we typically only have data in the $W1$ and $W2$ bands of WISE. We subtract the host contribution from the observations in order to measure the transient's flux, using the host contributions described above in Sect. \ref{detections}. 

We then estimated the BB parameters for the transients using the EMCEE python implementation of the Markov chain Monte Carlo (MCMC) method \citep{ForemanMackey2013} to fit the points and derive uncertainties. Examples of the fits and uncertainties are shown in Fig. \ref{fig:MCMC_examples}. The inferred temperatures and radii for all transients are shown in Fig. \ref{fig:BB_results}. Additionally, we show the BB parameters derived for Arp 299-B AT1 and AT 2017gbl from \citetalias{Mattila2018} and \citetalias{Kool2020}. These are generally consistent within the uncertainties with those found from our data, although the parameters derived from the final two observations of AT 2017gbl in our work diverge from those derived in \citetalias{Kool2020}. This may be due to our fit working less well when the BB becomes cooler, as the peak of the BB is not well sampled by our sparse data. As discussed in Sect. \ref{subsec:IRAS05189}, there were $J-$band data available for a single epoch of IRAS F05189$-$2524 and we performed BB fitting at that epoch with a 3 point SED. The resulting parameters are shown in Fig. \ref{fig:BB_results}, and are consistent with the results from the 2 point SED fits. Apart from some exceptions described below, the peak temperatures inferred from the BB fitting are consistent with hot dust and the evolution is a continuous decline (within the uncertainties) as expected for cooling dust. The radii inferred from these fits cannot be simply connected to a physical distance between the SMBH and the hot dust, as we do not expect a smooth spherical distribution of dust to have been present before the transient. Instead the distribution of dust available for heating will have been shaped by the presence of an AGN and be highly asymmetrical (see Sect. \ref{subsec:SED} below).

BB fitting of the data of ESO~286-IG019 yielded almost constant temperatures of $\sim$400 K. The peak of the BB at this temperature is at much longer wavelengths than the NEOWISE observations, so our fitting method cannot constrain the BB evolution. Furthermore, we have no NEOWISE pre-outburst observations for this transient and although the most recent two observations indicate the galaxy may have reached quiescence, further observations are required to be certain of this. Therefore, we conclude that the fit results are unreliable.

The BB fitting to the first epoch of observation for AM 0702-602N (not shown in Fig. \ref{fig:BB_results}) yields a high temperature of $2800_{-1200}^{+2300}$ K. The large uncertainties arise because the BB peak is at shorter wavelengths than the coverage of the NEOWISE data. If this emission is only related to the IR echo, then within the uncertainties its temperature can be consistent with the evaporation temperature for amorphous carbon grains \citep[$\sim$2000 K; see][ and references therein]{dwek2021}. An alternative explanation is that there is an additional hot SED component related to the fading SMBH flare at this epoch, as was the case for AT 2017gbl \citepalias{Kool2020}. In this case, the hot component has faded by the next epoch, where the temperature measurement is $1600_{-200}^{+400}$ K, which is then consistent with the expected evaporation temperature for silicate dust ($\sim$1500 K).
  
We calculate the luminosity associated with the BB of the transients using the Stefan-Boltzmann law applied to the inferred parameters. From this luminosity we estimate the total radiated energy from the transients by integrating the luminosity over time between the epochs. The results are shown in Fig. \ref{fig:BB_LE}.






\subsection{\label{subsec:SED}Spectral energy distribution fitting of host galaxies}

We have fitted the pre-outburst SEDs of the host galaxies of the transients in our sample using the CYprus models for Galaxies and their Nuclear Spectra  (CYGNUS) radiative transfer models\footnote{The models are publicly available at https://arc.euc.ac.cy/cygnus}, which include libraries for the emission from a spheroidal galaxy, starburst and the AGN torus predicted by the standard unified model. Additionally, a small library of polar dust models is used where the only free parameter is the temperature of the optically thick polar dust clouds assumed to be constant in the clouds. A review of the SED fitting methods is available in \citet{pereztorres2021}.

The spheroidal galaxy model is an evolution of the cirrus model of \cite{err03} and is described in more detail in \cite{efstathiou21a}. This model solves the radiative transfer in a spheroidal galaxy where the stars and dust are mixed and follow a S\'ersic profile with $n=4$. The model parameters are the optical depth of the spheroidal cloud, the e-folding time of the delayed exponential that defines the star formation history of the galaxy and the parameter $\psi^s$ that controls the intensity of starlight.

The starburst model is described in \cite{errs00} and \cite{es09} and has been used extensively for modelling the IR emission of both local and high-z starburst galaxies. The model treats a starburst as an ensemble of giant molecular clouds centrally illuminated by recently formed stars, incorporates the stellar population synthesis model of \citep{bc93, bc03}, and considers the effect of transiently heated grains or polycyclic aromatic hydrocarbons in the radiative transfer. The key parameters of the model are the initial optical depth of the molecular clouds, the age of the starburst and the e-folding time of the exponentially decaying star formation history.

The AGN model follows the prescription developed by \cite{err95} and \cite{efstathiou13} and can additionally include the effect of polar dust \citep{ehy95, efstathiou06, Mattila2018}. The model self-consistently solves the radiative transfer problem in the AGN torus to determine the temperature distribution of dust grains, which have a range of sizes and composition. The model also self-consistently takes into account absorption of emission from the hot inner torus by dust in the cooler outer parts of the torus depending on the inclination. As discussed in \cite{efstathiou06} and \cite{efstathiou21b}, an anisotropy correction factor needs to be applied to the `observed' AGN luminosity to account for the anisotropic emission of the torus, which is a general feature of all currently used torus models. The model of \cite{err95} treats the AGN disc as a tapered one, where the height of the disc increases with distance from the BH, but tapers off to a constant height in its outer part. In this work we used a large grid of tapered discs. The parameters of the model are the ratio of outer to inner radius $r_2/r_1$, the equatorial UV optical depth, the opening angle of the disc $\theta_{o}$, and the inclination $\theta_{i}$ (both measured from the pole).

\begin{figure}
   \centering
    \centering
        \resizebox{\hsize}{!}{\begin{tabular}{c}
        \includegraphics[angle = 0, origin = c, width=0.9\hsize]{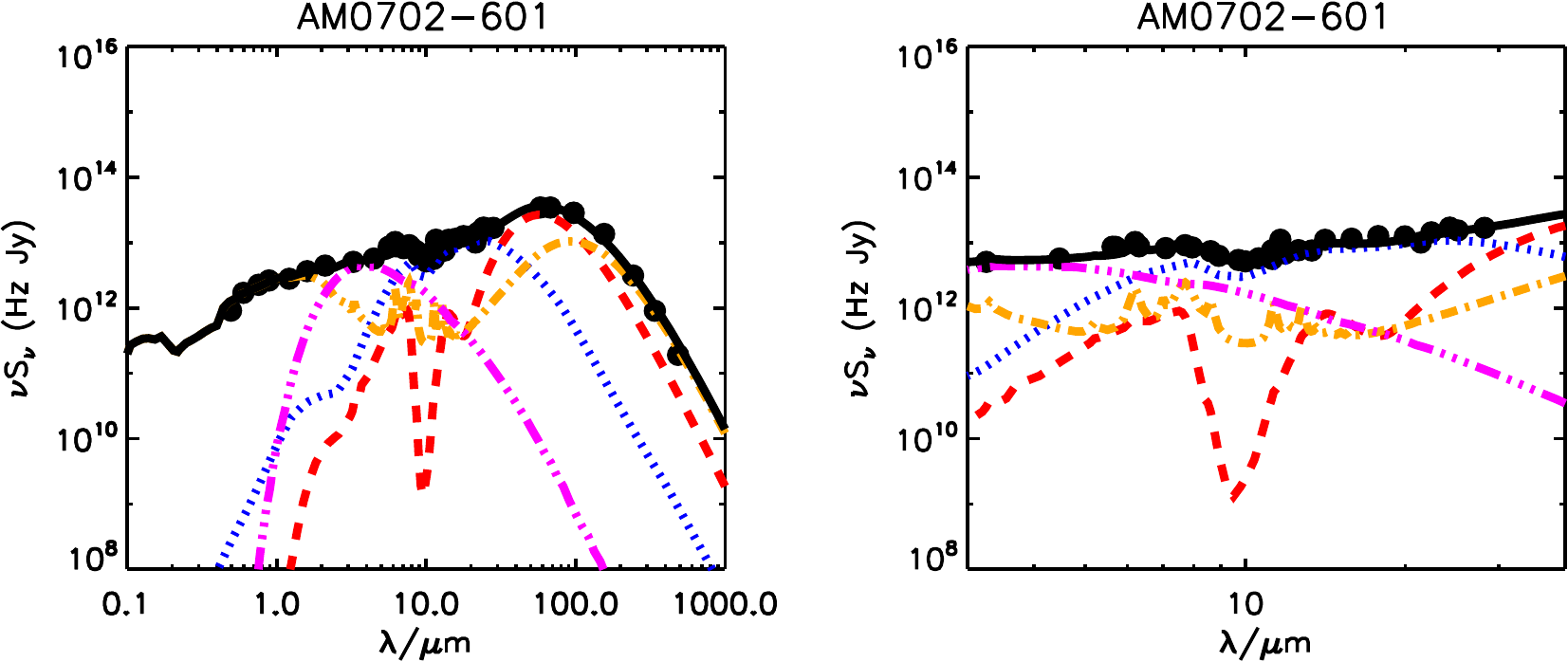} \\
        \includegraphics[angle = 0, origin = c, width=0.9\hsize]{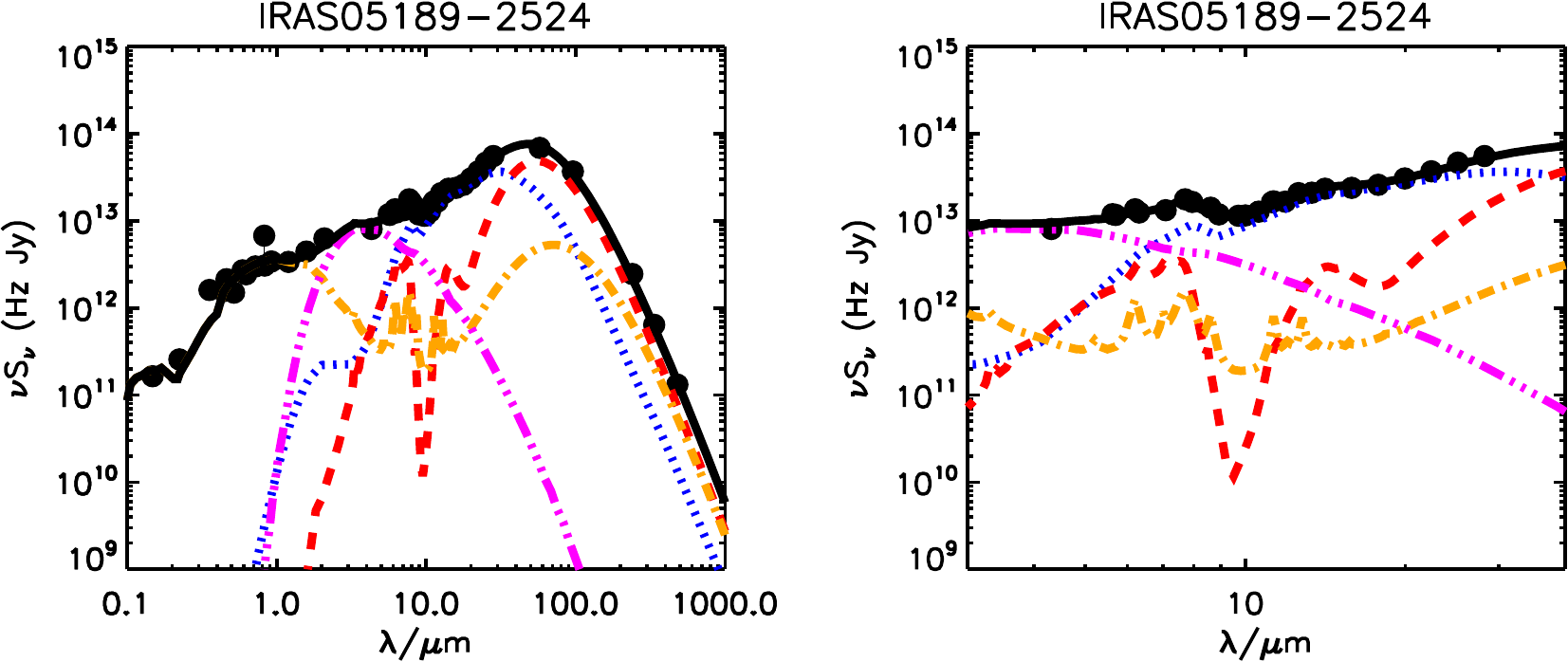} \\
        \includegraphics[angle = 0, origin = c, width=0.9\hsize]{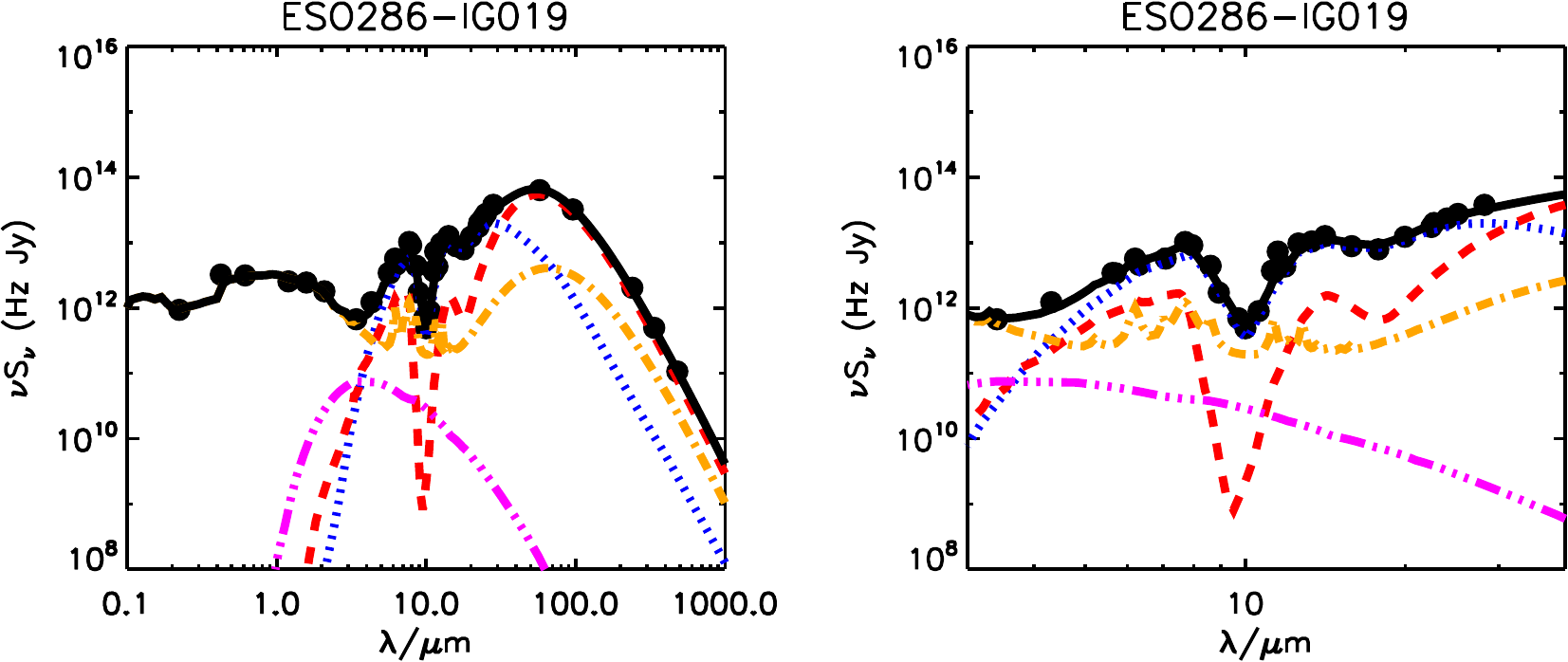}
        \end{tabular}}
         \caption{SED fits to the pre-outburst measurements of the host galaxies of the newly discovered transients in our sample. The left column shows the full wavelength range, and the right column shows a zoomed-in view of the region around 10 $\mu$m. Flux measurements are indicated by black dots. The model SED (black line) is composed of a starburst (red), an AGN torus (blue), a spheroidal host (orange), and a polar dust component (pink). Data sources are described in Sect. \ref{subsec:SED}.}
         \label{fig:SED_fits}
   \end{figure}

The data fitted in the case of IRAS~05189$-$2524 and ESO~286-IG019 are part of the compilation of multi-wavelength SEDs of the HERschel Ultra Luminous Infrared Galaxy Survey (HERUS) sample of galaxies presented by \citep{efstathiou21b}. The data for AM~0702-601N were assembled from publicly available archives such as CASSIS \citep{Lebouteiller2011} and NED. The spectral resolution of the Spitzer/IRS data was reduced so that they are better matched to the resolution of the radiative transfer models. To fit the SEDs, we used a grid of 7770 AGN torus templates, 3000 starburst templates, 780 spheroidal templates, and a single polar dust template for each SED that assumes a polar dust temperature of 900~K. The SEDs were fitted with the MCMC SED fitting code SATMC \citep{johnson13} and after post-processing of the fitting data the covering factor of polar dust $f_c$ and its uncertainty are computed. SATMC combines and interpolates these grids to generate the required number of models to reach convergence. The models and model parameters are presented in Fig. \ref{fig:SED_fits} and Table~\ref{tab:SED_fits}, respectively. We note that the modelling presented here for ESO 286-IG019 includes a polar dust component, in contrast to that presented in \citet{efstathiou21b}, in order to provide consistency across the modelling of our sample.

The fitting yields a large AGN contribution to the host galaxies of all the newly discovered transients, as well as a powerful starburst. It additionally supports the evidence for a Type 2 AGN in these galaxies also indicated by the WISE colours. For the angles $\theta_{i}$ and $\theta_{o}$ measured from the pole as we have defined them, the criteria for the central SMBH to be obscured is that $\theta_{i} - \theta_{o} > 0$, which is the case for all the hosts of our transients. Notably, the value of $\theta_{i} - \theta_{o}$ for ESO 286-IG019 is much larger than for the other two hosts, indicating a more edge-on view. This could lead to the polar dust being significantly obscured in this system, and this could be connected to the redder, less luminous transient that we observe in this galaxy. On the contrary, the relatively small values of $\theta_{i} - \theta_{o}$ for the other two systems indicate a quite clear view of the polar dust. Both AM 0702-601N and IRAS F05189$-$2524 have large contributions to their luminosity from polar dust, leading to significant covering factors, particularly in the case of AM 0702-601N. We discuss this fact in connection with potential interpretations of these transients in Sect. \ref{subsec:transientInterp}.

\begin{table*}
\caption{Parameters derived from SED fitting for the host galaxies of transients in our sample.}
\centering
\resizebox{\hsize}{!}{
\begin{tabular}{lcccccccccccc} \hline \hline
Galaxy & L$^{c}_{total}$ &  L$_{starburst}$ & L$^{o}_{AGN}$ &  L$^{c}_{AGN}$ & L$_{Polar}$ & Cov factor & AGN frac & $\theta_{i}$ & $\theta_{o}$ & CCSNe rate & SF rate & Ref \\ 
 & 10$^{11}$ L$_{\odot}$ &  10$^{11}$ L$_{\odot}$ & 10$^{11}$ L$_{\odot}$ &  10$^{11}$ L$_{\odot}$ & 10$^{9}$ L$_{\odot}$ & \% & \% & $^{\circ}$ &  $^{\circ}$ &  SN yr$^{-1}$ & M$_{\odot}$ yr$^{-1}$ &  \\ \hline
 
AT 2017gbl pre & 4.35$_{-0.08}^{+0.09}$ & 1.61$_{-0.10}^{+0.04}$ &  0.43$_{-0.02}^{+0.03}$ & 1.40$_{-0.09}^{+0.09}$ & 0.01$_{-0.006}^{+0.12}$  & 0.01$_{-0.01}^{+0.17}$ & 32.2$^{+2.2}_{-2.2}$ & 60.8$_{-0.7}^{+0.2}$ & 43.6$_{-1.0}^{+0.9}$ & $0.16^{+0.01}_{-0.01}$ & $15^{+2}_{-1}$  & (1,4) \\
AT 2017gbl post & 4.35$_{-0.08}^{+0.09}$ & 1.61$_{-0.10}^{+0.04}$ & 0.43$_{-0.02}^{+0.03}$ & 1.40$_{-0.09}^{+0.09}$ & 3.8$_{-0.2}^{+0.2}$  & -- & 32$^{+2}_{-2}$ & 60.8$_{-0.7}^{+0.2}$ & 43.6$_{-1.0}^{+0.9}$ & $0.16^{+0.01}_{-0.01}$ & $15^{+2}_{-1}$  & (1,4) \\ \hline

Arp 299-B AT1 pre & 1.30$_{-0.12}^{+0.14}$ & 0.98$_{-0.08}^{+0.12}$ & -- & 0.32$_{-0.09}^{+0.08}$ & 0.52$_{-0.52}^{+4.68}$ & 3$_{-3}^{+40}$ & 24$_{-6}^{+4}$ & -- & -- & 0.33 & 56  & (2),(3) \\
Arp 299-B AT1 post & 1.37$_{-0.08}^{+0.09}$ & 0.99$_{-0.02}^{+0.01}$ & -- & 0.30$_{-0.07}^{+0.13}$ & 7.78$_{-4.07}^{+4.05}$ & 52$_{-29}^{+26}$ & 23$_{-4}^{+7}$ & -- & -- & 0.33 & 56  & (2),(3) \\ \hline

AM 0702-601N & 7.3$_{-0.2}^{+0.2}$ & 1.76$_{-0.07}^{+0.07}$ & 0.93$_{-0.04}^{+0.09}$ & 4.3$_{-0.2}^{+0.3}$ & 35$_{-1}^{+2}$ & 16.2$_{-1.1}^{+0.8}$ & 59$_{-1}^{+2}$ & 64.2$_{-0.1}^{+0.8}$ & 54.0$_{-0.7}^{+0.3}$ & 0.37$_{-0.08}^{+0.06}$ & 28$_{-7}^{+4}$  & (4) \\ \hline

IRAS F05189$-$2524 & 56$_{-5}^{+6}$ & 7.7$_{-0.8}^{+0.7}$ & 5.5$_{-0.6}^{+0.8}$ & 47$_{-6}^{+6}$ & 98$_{-17}^{+15}$ & 4.2$_{-0.7}^{+0.7}$ & 84$_{-3}^{+2}$ & 68$_{-2}^{+1}$ & 60.0$_{-1.4}^{+0.4}$ & 0.9$_{-0.1}^{+0.3}$ & 70$_{-9}^{+21}$  & (5)\\ \hline

ESO 286-IG019 & 23$_{-2}^{+4}$ & 6.7$_{-0.3}^{+0.2}$ & 2.7$_{-0.2}^{+0.5}$ & 14$_{-2}^{+4}$ & 1.1$_{-0.7}^{+1.2}$ & 0.2$_{-0.1}^{+0.2}$ & 64$_{-3}^{+6}$ & 80.3$_{-1.4}^{+1}$ & 41.1$_{-0.3}^{+1.9}$ & 0.9$_{-0.2}^{+0.3}$ & 67$_{-14}^{+22}$  & (4)\\ \hline


\end{tabular}
}
\tablefoot{Luminosity values are bolometric. L$^{o}_{AGN}$ and L$^{c}_{AGN}$ denote the AGN luminosity pre- and post-correction for the anisotropy of the emission from the AGN torus, and L$^{c}_{total}$ includes this correction. $\theta_{i}$ denotes the inclination of the AGN torus, and $\theta_{o}$ is the torus opening angle, with both angles measured from the pole. The star formation rate of the starburst is averaged over the last 50 Myr. Values for Arp 299-B AT1 are for the Arp 299-B region and not for the entire galaxy system.}
\tablebib{(1) \citetalias{Kool2020}; (2) \citetalias{Mattila2018}; (3) \citet{Mattila2012}; (4) this work; (5) \citet{efstathiou21b}.}
\label{tab:SED_fits}
\end{table*}

\section{\label{sec:results}Results}

\subsection{\label{subsec:transientInterp}The nature of the transients}

Our sample of transients consists of three new discoveries and two previously studied transients. The previously observed transients, Arp 299-B AT1 and AT 2017gbl, were interpreted as IR echoes resulting from re-radiation of emission released from a flare most likely associated with a dust-obscured TDE by the central SMBH \citepalias{Mattila2018,Kool2020}. Additionally, this re-radiation was argued to originate in optically thick dust clouds in the polar regions of the host AGN, as the dusty torus would block any radiation along a direct line of sight to the central AGN regions, even in the IR. The BB temperatures of AM 0702-601N and IRAS F05189$-$2524 inferred by our fits show temperature evolution that is consistent with the IR echo interpretation, with the earliest temperatures being close to the dust evaporation temperature and then declining over time. Furthermore, the optical Type 2 /S1h AGN classification of the host galaxies of all our newly discovered transients, and our SED fitting, indicate an edge-on view of the dusty torus. It is therefore intuitive to also suggest a scenario in which the IR echo we observe is produced by polar dust. Our SED fitting reveals large values for the polar dust covering factor in AM 0702-601N and IRAS F05189$-$2824 supporting this scenario. In the case of IRAS 05189$-$2524, there is further evidence for this from observations of broad lines visible in optical polarised light \citep{Young1996}, which are produced by the scattering of the AGN optical light by the polar clouds \citep{Antonucci1985,Tran1992}. As we do not have coverage of the rise of the IR light curve for ESO 286-IG019, or a good measurement of the quiescent flux of the host, we cannot clearly say if this interpretation holds also for this object. However, we note that this object has a larger inclination of its dusty torus than the other transients, so our view to the polar dust may be more obscured by the torus, and this could lead to the less luminous and redder transient we observe in the two WISE bands.

The host galaxies of all the detected transients in our sample have significant contributions to their luminosities from AGN. Therefore it seems likely that the IR echoes we have observed are caused by optical flares related to events within the already existing AGN structures. As discussed in Sect. \ref{subsec:filtering}, our detection criterion attempted to remove galaxies that were undergoing `normal' AGN variability. However, we note that CLAGN have displayed mid-IR luminosity changes of a similar timescale and amplitude as those discussed here \citep{Sheng2017,Assef2018,Sheng2020}. CLAGN are accompanied by optical variability \citep{Macleod2016}, typically over 1000s of days \citep{Macleod2019}. We have poor direct observational constraints on the optical variability of our transients. For IRAS 05189$-$2524, we can infer that it was shorter than $\sim$1500d, as the entire transient evolution in $W1$ lasts less than that, and the mid-IR variability is expected to take place over a longer timescale than in the optical, due to light travel time effects \citep[see, for example,][]{2016MNRAS.458..575L}. For AM 0702-601N, there is evidence from the BB fitting that the first WISE point during the outburst has also some flux contribution from a hot BB, being likely due to a fading optical outburst. If so, as there is no evidence for any excess flux in the pre-outburst WISE observations it is possible that the hot (optical) component evolved quite rapidly. However, it is likely that any observable optical flux from these transients was at a low level relative to the IR, as in the cases of AT 2017gbl and Arp 299-B AT1, given the Type 2 viewing angle to these AGN implies that the dusty torus would obscure the direct line of sight to the central AGN regions.


An interesting possibility for the origin of these transients is a TDE. The energetics of AM 0702-601N and IRAS F05189$-$2524 are similar to those of Arp 299-B AT1 for which a TDE origin was found as the best explanation, due to the properties of the radio jet associated with it \citepalias{Mattila2018}. The light curve evolution of AM 0702-601N is surprisingly similar to that of Arp 299-B AT1, although it is slightly bluer. The estimated SMBH masses cover a range from log$_{10}$($M_{BH}/M_{\odot}$) = 7.1 for IRAS 23436+5257 \citepalias{Kool2020} and 7.3 for Arp 299-B \citep{ptak2015} to log$_{10}$($M_{BH}/M_{\odot}$) = 8.1 for AM 0702-601N and 7.5-8.6 for IRAS F05189$-$2524. A sharp drop in the rate of optical TDEs has been observed in galaxies hosting a SMBH more massive than log$_{10}$($M_{BH}/M_{\odot}$) = 7.5 \citep{vanvelzen2018}, which are expected to directly capture stars without a luminous flare if non-spinning. However, spinning SMBHs with masses up to log$_{10}$($M_{BH}/M_{\odot}$) = 8.8 have been found to be capable of disrupting solar-like stars \citep{kesden2012}. Also, in U/LIRGs undergoing a major merger stellar velocity dispersion based SMBH mass estimates might not be reliable and the strong nuclear obscuration makes BH mass determinations challenging.

Optically discovered TDEs exhibit some important selection biases towards dust free environments \citep{Jiang2021b} and galaxies that do not host an AGN \citep{vanVelzen2020}. Likely due to these biases, they have been shown to generally exhibit low luminosity IR echoes, with inferred covering factors of only 1\% \citep{Jiang2021b}, although there are some exceptions such as ASASSN-15lh \citep{Leloudas2016} and ATLAS17jrp \citep{Wang2022}. As shown in Table~\ref{tab:Parameters}, the IR echo from a typical optically discovered TDE such as ASSASN-14li is much less luminous and shorter lived than the IR echoes we observe \citep{2016MNRAS.458..575L,vanVelzen2016}, but objects such as ASASSN-15lh display IR echoes that are quite similar. If the transients discovered in this work also have their origin in TDEs, they can be expected to exhibit different properties from the optical sample due to interaction between the TDE debris and a pre-existing AGN accretion disc \citep{Chan2019}. Additionally, the large covering factors of polar dust that we show are present for the AGN hosting a number of our transients can naturally explain the much more luminous IR echoes they produce compared to the optically discovered TDEs. However, without more extensive multi-wavelength observations, it is difficult to securely identify our transients as TDEs.

Although all the host galaxies of the transients detected in our sample have high star formation rates, and implied core-collapse SN rates of 0.1--1 per year, we consider a SN origin to be unlikely. As discussed in Sect. \ref{subsec:filtering}, any transient passing our selection criteria is more luminous in the mid-IR than even the most luminous SNe observed. Furthermore, the total radiated energies of luminous interacting SNe such as SN 2010jl \citep{Fransson2014} are a factor of 3 lower than those of AT 2017gbl, the least energetic transient in our sample (see Table \ref{tab:Parameters}). However, there has been a SN scenario suggested for some nuclear transients with similarly high energies, such as PS1-10adi \citep{Kankare2017}, in which the authors suggest SN ejecta interacting with the dense medium in the nuclear environment could produce the large radiated energies \citep[see also][]{Grishin2021}. 

\subsection{\label{subsec:rates}Rates}

Considering the transients we observed in our sample, we can set a rate for such events in U/LIRGs. As we do not resolve each U/LIRG into its individual nuclei with the resolution of WISE, we consider each listed U/LIRG in the RBGS as a single galaxy for the purpose of this rate calculation. The NEOWISE observing period considered in this paper is December 2013 - December 2021 and each U/LIRG is observed with a $\sim$6~month cadence. We therefore take the monitoring time for each U/LIRG as 7.5 years, taking into account the average 3 months at the start and end of this time period over which a U/LIRG would not be monitored. There is a significant period of time before the beginning of the NEOWISE monitoring during which an outburst could occur and still be detected in our analysis, as is the case for Arp 299-B AT1 and ESO 286-IG019. The outburst in Arp 299-B AT1 occurred in December 2004, 9 years before the NEOWISE monitoring started. However, events such as AT 2017gbl and the transient in IRAS F05189$-$2524 would only have passed our luminosity constraint if they had occurred 1 and 3.5 years, respectively, before the monitoring period. 

To calculate the rate of such transients in U/LIRGs, we take our monitoring period plus the preceding 1 year as the sample period, as all of our detected transients would still have been detected had they begun to rise at that point. Our sample of 215 U/LIRGs observed across 7.5 + 1 years gives a 1827.5 year control time, in which five transients were detected. We assume a Poisson process and adopt the confidence limits of \cite{Gehrels1986} to obtain a rate for these transients of 10$^{n}$ U/LIRG$^{-1}$ year$^{-1}$ with $n = -2.56^{+0.22}_{-0.25}$ with 1$\sigma$ uncertainties, alternatively expressed as (1.6--4.6)$\times$10$^{-3}$ transients U/LIRG$^{-1}$ year$^{-1}$.


This rate is significantly higher than that observed by other surveys for nuclear transients. Another transient search in WISE data, described in \citet{Jiang2021a}, searched over a million galaxies for mid-IR outbursts, drawing their sample from those galaxies with redshift < 0.35 and available spectra from the Sloan Digital Sky Survey. They found an event rate of $\sim$5$\times$10$^{-5}$ galaxy$^{-1}$ year$^{-1}$, significantly lower than found in our survey of U/LIRGs and implying that there is an elevated rate of mid-IR outbursts within U/LIRGs compared to the general population of nearby galaxies. We note that the detection criterion they used is quite different from ours, depending on magnitude rather than luminosity changes, to the extent that only Arp 299-B AT1 and AT 2017gbl would pass the criterion of a 0.5 mag change in $W1$ or $W2$ that they impose. However, the most common peak luminosity of the host-subtracted transients they discovered is $\sim$10$^{43}$ erg s$^{-1}$, implying that their discoveries and ours do not draw from populations with significantly different energetics. 

Comparing to optical searches, \citet{Lawrence2016} estimate that 0.1\% to 0.01\% of AGN show `extreme' variability, which in their work refers to a change of at least 1.5 mag. We found five transients in our sample of 215 LIRGs, so that $\sim$2$\%$ of our sample are hosts to luminous transients. Optical searches for TDEs have found a rate of $\sim$10$^{-4}$ galaxy$^{-1}$ year$^{-1}$ \citep{vanvelzen2021,vanvelzen2018,Hung2018}. However, major mergers of galaxies, such as the LIRGs in our sample, are expected from simulations to have substantially higher TDE rates than the field galaxies, and post-starburst/E+A galaxies have also been observed to be over-represented as hosts of optically discovered TDEs \citep{Arcavi2014,French2020,Hammerstein2021}. Simulations show that major galaxy merger products can experience elevated TDE rates as the SMBHs can come into close proximity \citep{Li2019}. IRAS F05189$-$2524 and ESO 286-IG019 are both post-merger products that could be experiencing such a boost in the TDE rate. The separation of the nuclei, and the SMBHs they likely host, in AM 0702-601N and Arp 299 is still large, but the TDE rate could also be boosted by increased stellar densities in the nuclear regions due to the recent starburst episodes therein \citep{Stone2016}. We further note that the only optically discovered TDE candidate in a U/LIRG is hosted in IRAS F01004-2237, identified as having a dual AGN system based on SED fitting \citep{efstathiou21b}.


\section{\label{conclusions}Conclusions}
Data from the NEOWISE survey were used to search for IR luminous, smoothly evolving transients in a sample of 215 local LIRGs and ULIRGs. This effort was motivated by recent discoveries of energetic transients associated with TDEs at a higher than expected rate within LIRG and ULIRG host galaxies. As a result of the search, three new IR transients with luminosities exceeding $10^{43}$~erg s$^{-1}$ are reported here, and their properties investigated by means of BB fitting of the WISE measurements and radiative transfer modelling of their host galaxies. We note that all the host galaxies have a significant AGN contribution, indicating that a pre-existing accretion disc and dust surrounding the AGN probably play an important role in their occurrence. We find that the observed transient IR emission is consistent with IR echoes resulting from the re-radiation of shorter wavelength emission of transients associated with the SMBH at the centres of the host nuclei. Our detections of IR luminous transients correspond to a rate of (1.6-4.6)$\times$10$^{-3}$ / yr / galaxy, which is over an order of magnitude larger than the rate of occurrence of similarly luminous outbursts observed in AGN in the optical. This rate is also significantly higher than the rate of optically discovered TDEs in E+A galaxies. Such an enhanced TDE rate in galaxy mergers with increased stellar densities within their nuclear regions is not unexpected and has been suggested by simulations. If such events are common within LIRGs, there are implications for our understanding of galaxy evolution through the TDE contribution to the accretion of their SMBHs and possible feedback effects. Furthermore, TDEs occurring within AGN can be expected to exhibit properties different from those of the optical sample due to interaction between the TDE debris and a pre-existing accretion disc. However, other possible interpretations for these transients can also exist, such as an extremely energetic SN occurring within the dense nuclear environment of an AGN or some form of CLAGN scenario.

In the case of the transient in AM 0702-601N, an origin in a TDE is supported by our radio observations. A secure identification of these transients as TDEs may, however, only be feasible with the help of more extensive multi-wavelength observations such as the high-resolution radio observations presented in \citetalias{Mattila2018} in the case of Arp 299-B AT1. Additionally, future instruments such as the Son of X-shooter (SoXS; \citet{Schipani2016}) and the NOT Transient Explorer (NTE) will be able to provide medium-resolution spectra simultaneously in the optical and near-IR regions, the latter of which is less affected by dust extinction. In order to identify highly obscured transients early enough to obtain this high quality data of their early evolution, searches and surveys in the IR with a higher cadence and more rapid data processing than that offered by the NEOWISE survey are required. The future Roman Space Telescope will be critical in this respect.

\begin{acknowledgements}
We would like to thank Suvi Gezari and Takashi Nagao for useful discussion about the manuscript draft. We also thank the anonymous referee for very useful comments that led to substantial improvement to the manuscript after the initial submission. TMR acknowledges the financial support of the Jenny and Antti Wihuri foundation and the Vilho, Yrj{\"o} and Kalle V{\"a}is{\"a}l{\"a} Foundation of the Finnish academy of Science and Letters. ECK acknowledges support from the G.R.E.A.T research environment funded by {\em Vetenskapsr\aa det}, the Swedish Research Council, under project number 2016-06012, and support from The Wenner-Gren Foundations.

This publication makes use of data products from the Wide-field Infrared Survey Explorer, which is a joint project of the University of California, Los Angeles, and the Jet Propulsion Laboratory/California Institute of Technology, and NEOWISE, which is a project of the Jet Propulsion Laboratory/California Institute of Technology. WISE and NEOWISE are funded by the National Aeronautics and Space Administration.

This research has made use of the NASA/IPAC Infrared Science Archive, which is funded by the National Aeronautics and Space Administration and operated by the California Institute of Technology.

This work is partially based on observations obtained as part of the VISTA Hemisphere Survey, ESO Progam, 179.A-2010 (PI: McMahon).

This work is based in part on observations made with the Nordic Optical Telescope, owned in collaboration by the University of Turku and Aarhus University, and operated jointly by Aarhus University, the University of Turku and the University of Oslo, representing Denmark, Finland and Norway, the University of Iceland and Stockholm University at the Observatorio del Roque de los Muchachos, La Palma, Spain, of the Instituto de Astrofisica de Canarias.

This work is partly based on the NUTS2 programme carried out at the NOT. NUTS2 is funded in part by the Instrument Center for Danish Astrophysics (IDA).

The Australia Telescope Compact Array is part of the Australia Telescope National Facility (grid.421683.a) which is funded by the Australian Government for operation as a National Facility managed by CSIRO. We acknowledge the Gomeroi people as the traditional owners of the Observatory site.

The Australian SKA Pathfinder is part of the Australia Telescope National Facility (grid.421683.a) which is managed by CSIRO. Operation of ASKAP is funded by the Australian Government with support from the National Collaborative Research Infrastructure Strategy. ASKAP uses the resources of the Pawsey Supercomputing Centre. Establishment of ASKAP, the Murchison Radio-astronomy Observatory and the Pawsey Supercomputing Centre are initiatives of the Australian Government, with support from the Government of Western Australia and the Science and Industry Endowment Fund. We acknowledge the Wajarri Yamatji people as the traditional owners of the Observatory site.

This research made use of Astropy,\footnote{http://www.astropy.org} a community-developed core Python package for Astronomy \citep{Astropy2013,Astropy2018}. 

\end{acknowledgements}

\bibliographystyle{aa} 
\bibliography{bibliography} 

\begin{appendix}
\onecolumn
\section{Additional tables and figures}

\begin{figure}[h]
\centering
   \includegraphics[width=\textwidth]{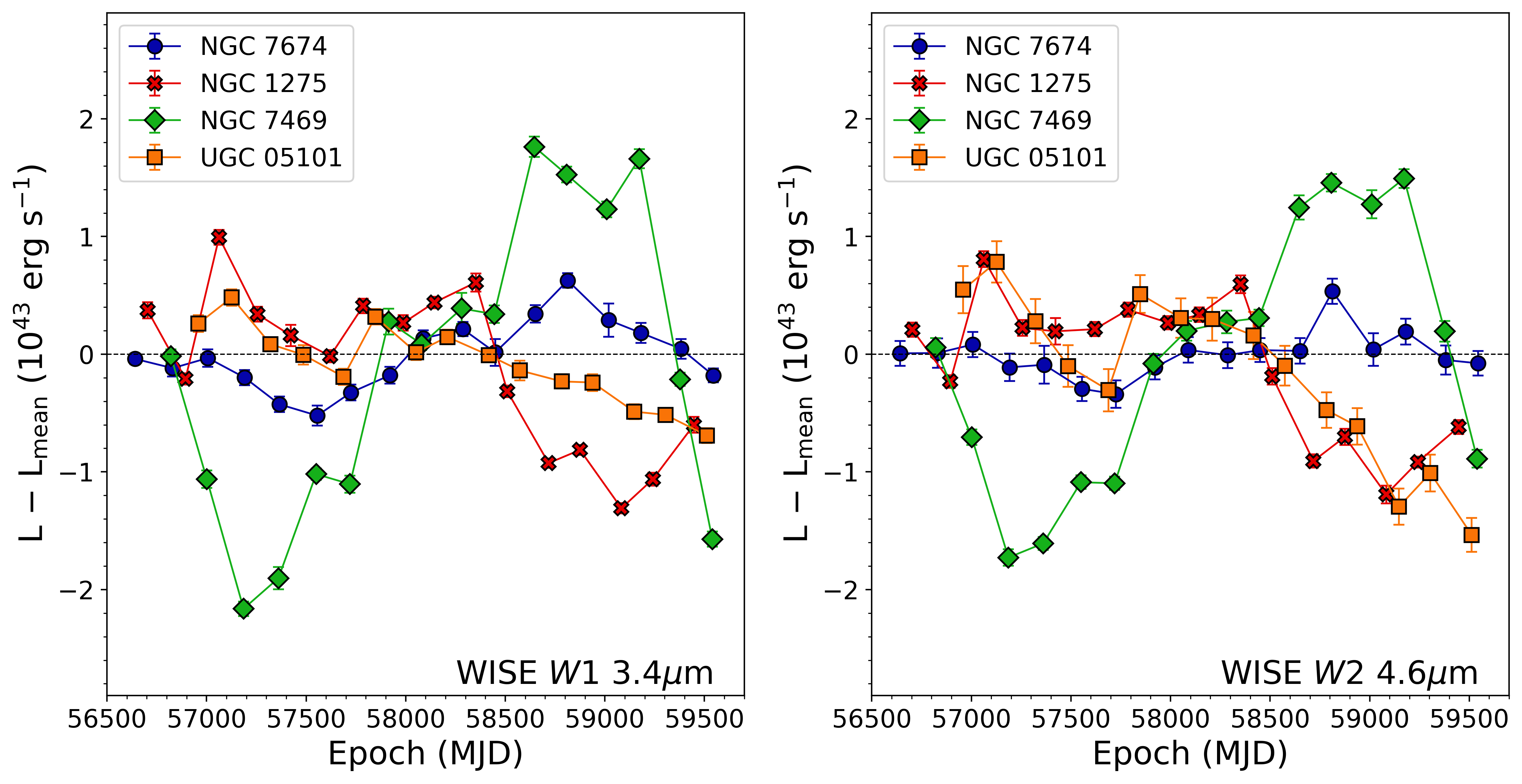}
     \caption{Luminosity evolution in WISE $W1$ and $W2$ bands of the rejected candidates identified by our transient detection process. The luminosity subtracted was the mean of all the luminosity measurements.}
     \label{fig:Not Transients}
\end{figure}   

\begin{figure}[h]
   \centering
   \includegraphics[width=0.48\textwidth]{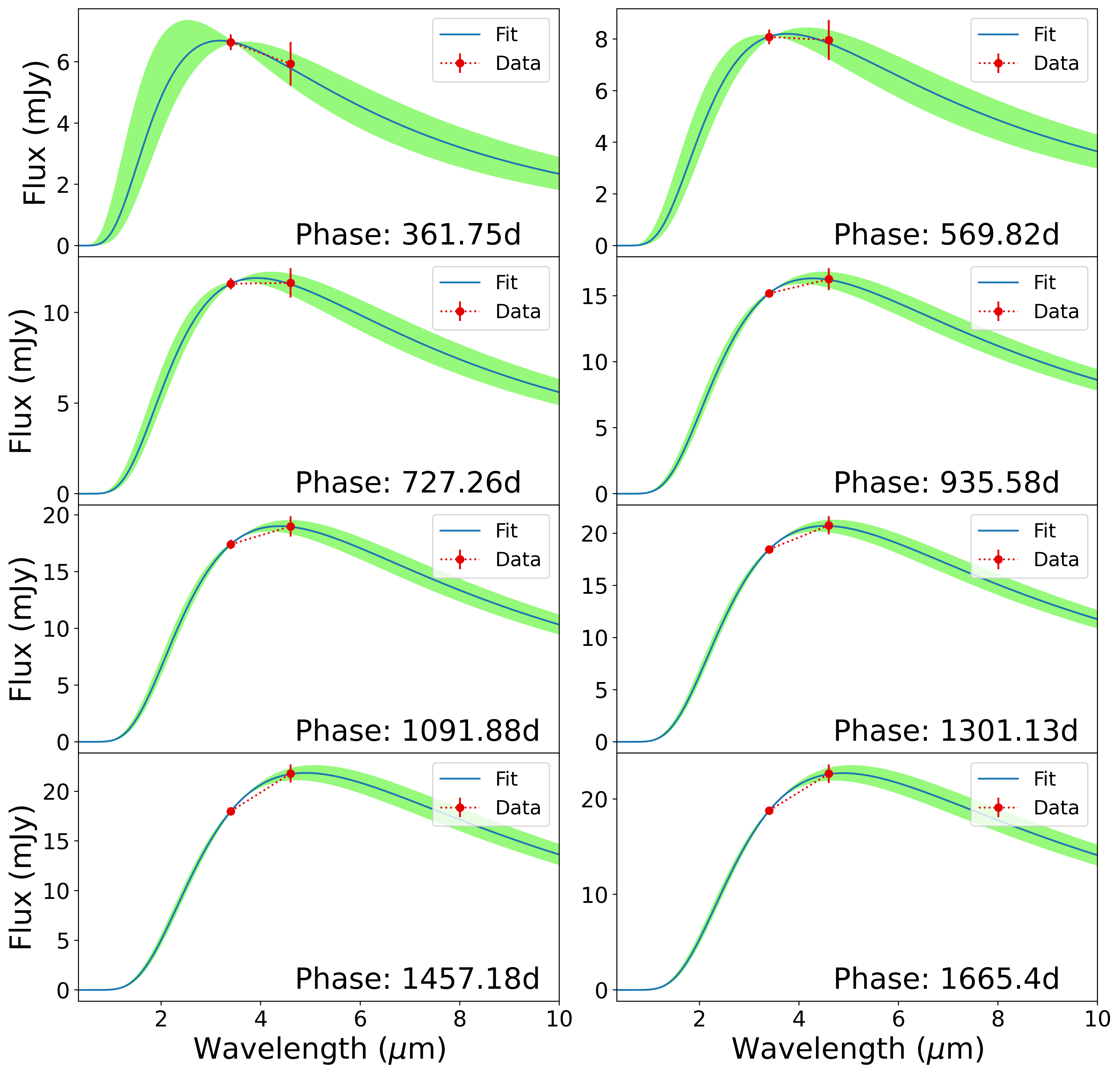} \quad
   \includegraphics[width=0.48\textwidth]{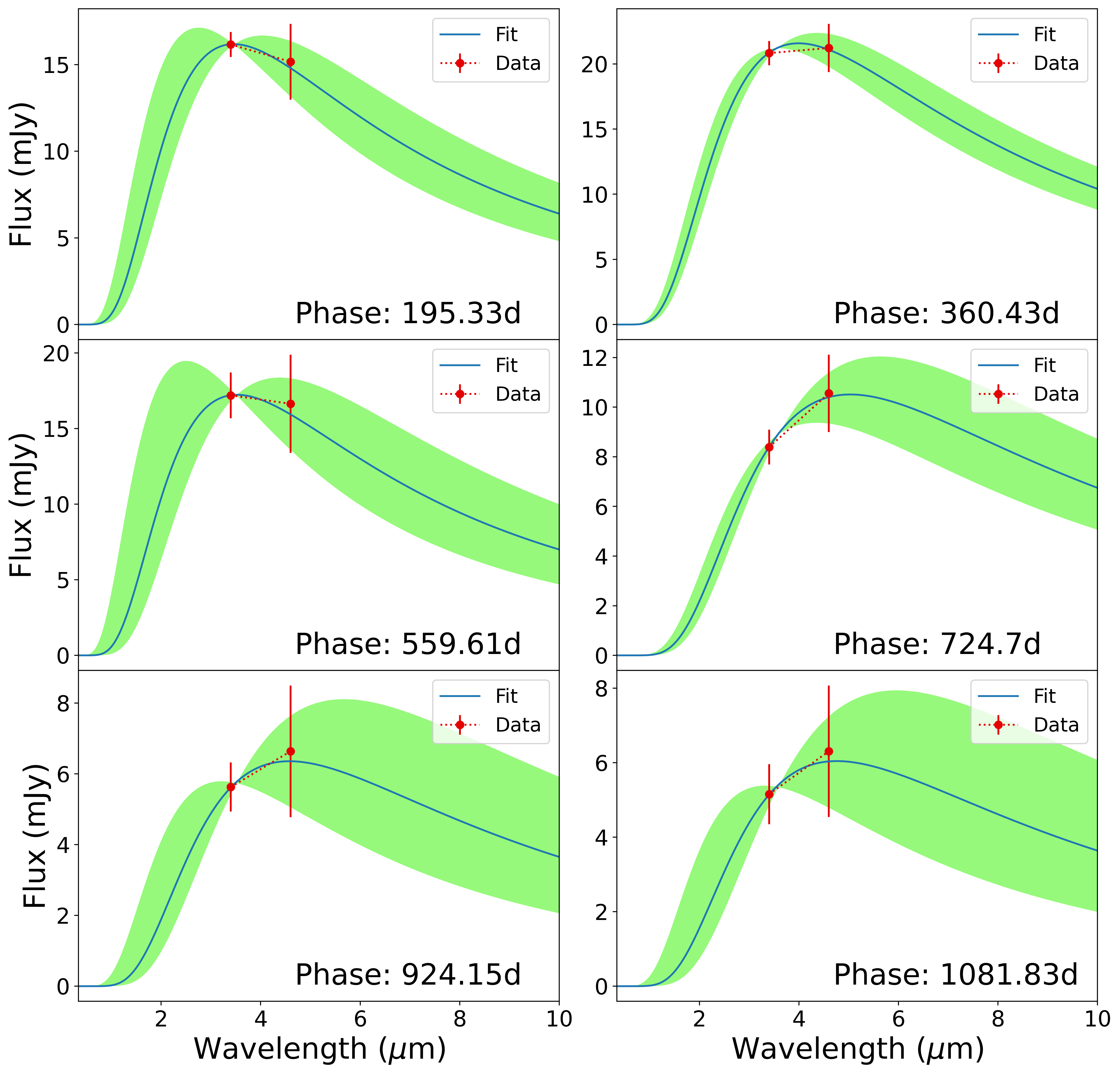}
      \caption{BB fits to AM 0702-601N and IRAS 05189$-$2524. The shaded regions indicate the 1 sigma uncertainties, measured as the 16th and 84th percentile values taken from the MCMC walkers.}
         \label{fig:MCMC_examples}
\end{figure}


\begin{longtable}{lccccccc}
\caption{Sample of U/LIRGs used for the analysis.}\\ 
\hline\hline
RBGS Name       &        AllWISE name   &        RA     &        Dec    &        Distance        &    L$_{IR}$ \\
        &               &        deg    &        deg    &        Mpc    &        log(L$_{\odot}$)        \\ \hline
 \endfirsthead
 \caption{continued} \\ 
 \hline
 RBGS Name      &        AllWISE name   &        RA     &        Dec    &        Distance        &    L$_{IR}$ \\
        &               &        deg    &        deg    &        Mpc    &        log(L$_{\odot}$)        \\ \hline
        \endhead
        \hline
        \endfoot
        \caption[]{The RBGS name for a galaxy being the same in multiple rows indicates that multiple WISE sources are associated with that galaxy, and the AllWISE names can be used to identify these sources. As described in the text, the distances are obtained from the NED and the value of L$_{IR}$ is updated from that given in the RBGS. The full table is available at the CDS.}
        \endlastfoot
NGC 0023        &         J000953.41+255526.0   &         2.4726        &         25.9239        &         64.4          &         11.12         \\
NGC 0034        &         J001106.54$-$120627.5         &         2.7773         &         $-$12.1076    &         82.8          &         11.50          \\
MCG-02-01-051/2         &         J001850.89$-$102236.8         &         4.7121         &         $-$10.3769    &         115.0         &         11.48          \\
MCG-02-01-051/2         &         J001849.98$-$102140.1         &         4.7083         &         $-$10.3612    &         115.0         &         11.48          \\
ESO 350-IG038   &         J003652.45$-$333316.9         &         9.2186         &         $-$33.5547    &         87.6          &         11.29          \\
NGC 0232        &         J004245.81$-$233341.0         &         10.6909         &         $-$23.5614    &         95.5          &         11.46          \\
... \\         &               &               &               &                &               \\

\label{tab:sample}

\end{longtable}

\end{appendix}

\end{document}